\documentclass[pra,onecolumn,superscriptaddress,aps,5pt]{revtex4}
\usepackage{amsfonts}
\usepackage{amssymb,latexsym,amsmath}
\usepackage{graphicx}
\usepackage{float}
\usepackage{dcolumn}
\usepackage{times}
\usepackage{subfigure}
\usepackage[toc,page,title,titletoc,header]{appendix}
\usepackage[usenames]{color}
\usepackage{ulem}

\newcommand{\sinc}{\mathrm{sinc}}
\newcommand{\sgn}{\mathrm{sgn}}

\begin{document}

\title{Traveling Waves and their Tails in Locally Resonant
Granular Systems}

\author{H. Xu}
\affiliation{Department of Mathematics and Statistics, University of
Massachusetts,
Amherst MA 01003-4515, USA}

\author{P.G. Kevrekidis}
\affiliation{Department of Mathematics and Statistics, University of
Massachusetts,
Amherst MA 01003-4515, USA}

\affiliation{Center for Nonlinear Studies and Theoretical Division, Los Alamos
National Laboratory, Los Alamos, NM 87544}

\author{A. Stefanov}
\affiliation{Department of Mathematics
University of Kansas
1460 Jayhawk Blvd Lawrence, KS 66045--7523 }

\begin{abstract}
In the present study, we revisit the theme of wave propagation
in locally resonant granular crystal systems, also
referred to as Mass-in-Mass systems. We use 3 distinct
approaches to identify relevant traveling waves. The first
 consists of a direct solution of the traveling
wave problem. The second one consists of the solution of the
Fourier tranformed variant of the problem, or, more precisely, of its convolution reformulation
(upon an inverse Fourier transform) of the problem in
real space.
Finally, our third approach will restrict considerations
to a finite domain, utilizing the notion of Fourier
series for important technical reasons, namely the
avoidance of resonances, that will be discussed in detail.
All three approaches can be utilized in
either the displacement or the strain formulation.
Typical resulting computations in finite domains
result in the solitary waves bearing symmetric
non-vanishing tails at both ends of the computational
domain. Importantly, however, a countably infinite set of
resonance conditions is identified for which solutions with genuinely
monotonic decaying tails arise.
\end{abstract}

\maketitle

\section{Introduction}

The theme of granular chains has gained considerable traction
over the past decade, perhaps to a large measure due to
the significant advances in corresponding experimental techniques
complementing theoretical and numerical investigations~\cite{Nesterenko2001,Sen2008,pgk_review,Theocharis_rev}. While traveling waves in these
extensively tunable (regarding, e.g., their homogeneous or heterogeneous,
weakly or strongly nonlinear nature) media have received the lion's share
of the corresponding attention, numerous other excitations have
been examined more recently as well including, but not limited to,
defect modes,
bright and dark breathers, and shock waves, among others;
see e.g.~\cite{Daraio2006bb,Hong2005,Fraternali2008,Doney2006,dev08,Coste2008,Daraio2004,Daraio2006,deBilly2000,DeBilly2006,Gilles2001,Gilles2003,Nesterenko1983,Porter2008,Porter2009,Rosas2004,Shukla1991,Molinari2009,Herbold2007,Daraio2006b,Herbold2007b,Hong2002,Job2005,Job2007,Melo2006,Boechler2010,Job2009,Theocharis2009,chong2013,sen3,sen4,vakakis1,vakakis2,vakakis3,stefanov1,stefanov2,dcdsa,Leonard13,pego1,chatter,pikovsky} for some characteristic examples.
A key feature supporting this explosive rate of development
has been the versatile use of such settings in a diverse
host of applications including e.g., actuating devices \cite{dev08}, acoustic lenses~\cite{Spadoni}, mechanical diodes~\cite{china1,china2,Nature11},
logic gates~\cite{Feng_transistor2014} and sound scramblers \cite{dar05,Nesterenko2005}.

The principal theme of robust, highly localized traveling stress
waves in such a setting has received arguably the most attention,
as it has posed significant mathematical challenges and has proved
fruitful in a diverse array of applications including, among others,
e.g. the quantification of bone-quality in biomedical applications~\cite{jk1}
or
the monitoring of hydration of gypsum cement~\cite{jk2}. From a
theoretical perspective, the existence of such waves,
originally established
in~\cite{Nesterenko1983} (see also~\cite{Nesterenko2001}),
 was rigorously
proved  in~\cite{ref7}
by means of the variational approach of~\cite{ref9}.
However, the non-constructive nature of this proof offered no information
on the solutions' profile. Their single pulse character (in the strain
variables of the system) was subsequently established
rigorously in~\cite{stefanov1}, following the
approach of~\cite{pego1}. These approaches also enabled
the identification of the doubly exponential character of
their spatial decay in the absence of the so-called precompression
(i.e., of linear part within the system dynamics); see also
for earlier relevant asymptotic analyses predicting
this rate of spatial decay~\cite{chatter,pikovsky}.

While this part of the story is admittedly well-understood in the
context of the standard one-component, homogeneous granular chain,
 in recent years, a number
of intriguing variants have emerged on this theme which may possess
more elaborate structural characteristics as regards their traveling
waves (and other related structures). Chronologically, an earlier
example of this form consists of the so-called cradle system
(of which we are not aware of an experimental realization as of yet)
proposed in~\cite{ref28} and further explored, including by detailed
direct numerical computations in~\cite{ref29}. There, a local,
linear oscillator was added to the granular chain, emulating
the pendulum restoring effect in the well-known Newton's cradle.
In that context, numerous unexpected features arose including
the formation of waves with persistent tails, of traveling (time-periodic)
breathers, of apparently direction-reversing (so-called boomeron-type)
structures, etc. While many of these observations have yet to be
explained, here we will actually turn our attention to another
class of systems, the so-called locally resonant granular crystals,
otherwise known as mass-in-mass (MiM) or mass-with-mass (MwM) systems.

The MiM and MwM systems are rapidly gaining an increasing amount of
interest chiefly because they have already been experimentally
realized in~\cite{luca} and~\cite{gatz}, respectively. Admittedly,
both of these realizations were chiefly linear (in the presence
of externally imposed pre-compression of the chain)
and aiming to illustrate the remarkable
meta-material type properties that these systems possess. Yet,
while a MwM nonlinear prototype was also demonstrated in~\cite{annavain},
it was a different type of experiment that very recently realized
highly nonlinear propagation in a locally resonant granular
system~\cite{yang}. In particular, the experiment of~\cite{yang}
built a so-called woodpile configuration consisting of
orthogonally-stacked  rods (every second rod is aligned, in this
alternating 0-90 degree configuration) and demonstrated that
the internal vibrations of the rods can play the role of
the local resonator within the granular chain.
It was also shown that depending on the properties of the system
(i.e., the length of the rods), one can controllably incorporate
one or more such resonators, yet here our focus will be in the
case of a single such resonator.

A remarkable find of these locally resonant, highly nonlinear
such chains was found to be the propagation of traveling waves
with an apparently non-decaying tail~\cite{yang}. It was indeed
found that the tail behind the wave may persist in small amplitude
oscillations (in the experiment they were found to potentially
be three order of magnitude smaller than the core of the
traveling wave) which, however, do not decay and which bear
a clear frequency/wavenumber, namely that of the of out-of-phase
vibration between the principal and the resonating mass
in such a system. Weakly nonlocal solitary waves have been studied
extensively in a series of examples in physical sciences and
engineering~\cite{remois,boyd}, yet very few (especially so,
highly experimentally controllable) realizations thereof are
known to exist. Our aim in the present work is to examine
more systematically such traveling waves from a theoretical
perspective.

We will use three distinct approaches in order to identify this
important class of solutions for the MiM/MwM systems. We will
seek them directly as traveling waves in real space, by attempting
to identify
fixed points of a discretization of the co-traveling frame
nonlinear problem in our direct method. We will also
follow the approach of~\cite{pego1}, rewriting the problem
in Fourier space (upon a Fourier transform) and seeking fixed
points of that variant
upon inverse Fourier transform, similarly to~\cite{pego1,stefanov1}.
This will recover a convolution based reformulation of the problem
in real space. Finally, the third method will recognize the limitations
of the infinite domain formulation of the above convolution problem
and will instead restrict consideration to a finite lattice, using
Fourier series rather than Fourier trasnforms.
The presentation of the three
methods will take place in section II below.
These methods will be explored
in a {\it finite domain} setting
yielding convergence to exact
traveling waves under suitable conditions that will be explicitly analyzed.
In addition to the, arguably more tractable, pulse-like solutions of
the so-called strain variant of the problem, we will also
translate the results at the level of bead displacements.
Relevant
numerical results will be given in section III of the manuscript,
along with some of the nontrivial associated challenges of the
computations and impact of parametric variations on the resulting
waves. In that process, a particularly interesting feature will
be identified (and subsequently explained), namely for an isolated,
countably infinite set of parameters the problem will no
longer possess the oscillating tails described above, but
rather the symmetric, monotonically decaying tails
of the standard granular chain. Finally, in section IV,
we will summarize our findings and present some potential directions
for future work.

\section{Theoretical setup}

\subsection{Model and Traveling Wave Formulation}

Our starting point will consist of a Hertzian chain~\cite{Nesterenko2001}
of identical beads with the displacement of the $n^{th}$ bead from the original
position denoted by $u_n$. For the $n^{th}$ bead $(n=1,2,...,N)$,
we attach a local resonator, effectively coupling it to another kind of bead
(the ``in-mass''~\cite{luca} or the ``with-mass''~\cite{gatz}
discussed above), whose displacement from the original position is
denoted by $v_n$. Notice that in the case of the woodpile configuration,
this does not constitute a separate mass but rather reflects the internal
vibrational modes of the woodpile rods~\cite{yang}.
Here we use $k_1$ to stand for the spring constant and $\nu_1$ reflects the ratio of two kinds of masses (or more generally, the effective mass of the
locally resonant mode).
With these assumptions, we write our model of the MiM/MwM problem as follows:
\begin{eqnarray}
\label{eqn1_1}
\ddot{u}_n&=&(\delta_0+u_{n-1}-u_n)_+^p-(\delta_0+u_{n}-u_{n+1})_+^p-k_1(u_n-v_n),\\
\label{eqn1_2}
\nu_1\ddot{v}_n&=&-k_1(v_n-u_n).
\end{eqnarray}
Here $n\in\mathbf{Z}$, and $\delta_0$ is the initial pre-compression between
beads, while $p=3/2$ is determined by the Hertzian interaction.

Focusing on traveling wave solutions of the above system, we
seek them in the form $u_n(t)=r(n-ct)=r(\xi)$ and $v_n(t)=s(n-ct)=s(\xi)$,
i.e., going to the co-traveling frame with $c$ as the traveling speed,
obtaining advance-delay differential equations for the profile
dependence on the
co-traveling frame variable $\xi$:
\begin{eqnarray}
\label{eqn2_1}
c^2\ddot{r}(\xi)&=&(\delta_0+r(\xi-1)-r(\xi))_+^p-(\delta_0+r(\xi)-r(\xi+1))_+^p-k_1(r(\xi)-s(\xi)),\\
\label{eqn2_2}
c^2\nu_1\ddot{s}(\xi)&=&-k_1(s(\xi)-r(\xi)).
\end{eqnarray}
Moreover, if we consider the equations of relative displacements
(i.e., the strain variables), writing $R(\xi)=r(\xi-1)-r(\xi)$ and $S(\xi)=s(\xi-1)-s(\xi)$, the equations assume the form:
\begin{eqnarray}
\label{eqn3_1}
c^2\ddot{R}(\xi)&=&(\delta_0+R(\xi+1))_+^p+(\delta_0+R(\xi-1))_+^p-2(\delta_0+R(\xi))_+^p-k_1(R(\xi)-S(\xi)),\\
\label{eqn3_2}
c^2\nu_1\ddot{S}(\xi)&=&-k_1(S(\xi)-R(\xi)).
\end{eqnarray}
The discrete version of the strain equation can be obtained as:
\begin{eqnarray}
\label{eqn3_01}
\ddot{x}_n(t)&=&(\delta_0+x_{n+1})_+^p+(\delta_0+x_{n-1})_+^p-2(\delta_0+x_n)_+^p-k_1(x_n-y_n),\\
\label{eqn3_02}
\nu_1\ddot{y}_n(t)&=&-k_1(y_n-x_n).
\end{eqnarray}
where $x_n(t)=u_{n-1}-u_n$ and $y_n(t)=v_{n-1}-v_n$.

It should be mentioned that if $r$ and $s$ are solutions to Eqs.~(\ref{eqn2_1})-(\ref{eqn2_2}) with current parameters, $\tilde{r}=a^4 r$ and $\tilde{s}=a^4 s$ will solve those equations with $\tilde{c}=ac$, $\tilde{k_1}=a^2 k_1$ and $\tilde{\delta_0}=a^4 \delta_0$. Similarly, there is a family of solutions to Eqs.~(\ref{eqn3_1})-(\ref{eqn3_2}) with different speeds, amplitudes and other parameters.

Our first solution method of this
system will consist of a direct method  attempt at
solving this problem at the level of Eqs.~(\ref{eqn2_1})-(\ref{eqn2_2})
for the displacements or of Eqs.~(\ref{eqn3_1})-(\ref{eqn3_2})
for the strains in real space. This approach is denoted by ``scheme I"
in what follows.

\subsection{Infinite domain with Fourier Transform}

In this subsection, we consider the infinite domain situation
where $\xi\in(-\infty, \infty)$, bearing in mind, however, that
in realistic setups this is practically irrelevant, as all
numerical computations and experimental observations are
conducted in finite domain settings. Additionally, our analysis
will be restricted to the highly nonlinear regime whereby $\delta_0=0$.

Here we assume $R$ and $S$ are functions such that Fourier transform $\hat{f}(k)=\int_{-\infty}^{\infty}f(\xi)e^{-ik\xi}d\xi$ can be applied on $R$, $S$, $\ddot{R}$, $\ddot{S}$ and $(R^{3/2})_+$. This condition is necessary because the above-mentioned experimental observations~\cite{yang}
suggest that $R(\xi)$ and $S(\xi)$ sometimes bear non-decaying (yet bounded)
oscillating tails as $|\xi|\to\infty$. This implies that
 $R$ and $S$ may be non-integrable and thus their Fourier transforms
may not be possible to define. In this subsection we will only focus
on the special cases for $R$ and $S$ where such
integrable solutions do exist.

With the assumption above,
we apply Fourier Transform to Eqs.~(\ref{eqn3_1})-(\ref{eqn3_2}) to obtain:
\begin{eqnarray}
\label{eqn4_1}
(k_1+k_1\nu_1-c^2\nu_1 k^2)\hat{R}&=&\frac{1}{c^2}(k_1-c^2\nu_1 k^2)\sinc^2{(\frac{k}{2})}\widehat{(R^{3/2})_+},\\
\label{eqn4_2}
(k_1-c^2\nu_1 k^2)\hat{S}&=&k_1\hat{R}
\end{eqnarray}with $\sinc(k)$ being defined as $\frac{\sin(k)}{k}$.

In order to make sure that $\hat{R}$, $\hat{S}$ and $\widehat{(R^{3/2})_+}$ are well-defined functions, extra conditions should be imposed:

(a). $\hat{R}(\pm\sqrt{\frac{k_1}{c^2\nu_1}}) \equiv\hat{R}(\pm k_2)=0$ ;

(b). $\widehat{(R^{3/2})_+}(\pm\sqrt{\frac{k_1(1+\nu_1)}{c^2\nu_1}})
\equiv \widehat{(R^{3/2})_+}(\pm k_0)=0$ or $k_0=2 n \pi$.

Here we defined $k_0=\sqrt{\frac{k_1(1+\nu_1)}{c^2\nu_1}}$ and $k_2=\sqrt{\frac{k_1}{c^2\nu_1}}$.

If we choose $k_0=2 n \pi$, the condition (b) is automatically
met and the zeros of $\hat{R}$ will be investigated later for condition (a). In this case,
the second one among Eqs.~(\ref{eqn4_1})-(\ref{eqn4_2}) can be directly solved and
back-substituted in the first, to yield $\hat{R}=\frac{k_1-c^2\nu_1 k^2}{c^2(k_1+k_1\nu_1-c^2\nu_1 k^2)}\sinc^2{(\frac{k}{2})}\hat{(R^{3/2})_+}
\equiv M_1(k)\hat{(R^{3/2})_+}$. Since $\lim_{k\to\pm k_0}\frac{ {\rm sinc}^2(\frac{k}{2})}{k^2-k_0^2}=0$, the singularities of $M_1(k)$ at $\pm k_0$ can be removed. Moreover, if we define the Inverse Fourier Transform of $\hat{f}(k)$ as $f(\xi)=\frac{1}{2\pi}\int_{-\infty}^{\infty}\hat{f(k)}e^{i k \xi} dk$, then the inverse Fourier transform of $M_1(k)$ is
\begin{equation}
\begin{split}
\label{eqn_m1}
m_1(\xi)=&\mathcal{F}^{-1}[\frac{1}{c^2}(1+\frac{k_1}{c^2}\frac{1}{k^2-k_0^2})\sinc^2(\frac{k}{2})]\\
=&\frac{1}{c^2}\mathcal{F}^{-1}[\frac{\sin^2(\frac{k}{2})}{(\frac{k}{2})^2}(1-\frac{k_1}{c^2 k_0^2})+\frac{k_1}{c^2 k_0^2}\frac{\sin^2(\frac{k}{2})}{(\frac{k^2-k_0^2}{4})}]\\
=&\frac{1}{c^2}(1-\frac{k_1}{c^2 k_0^2})\mathcal{F}^{-1}[\frac{\sin^2(\frac{k}{2})}{(\frac{k}{2})^2}] + \frac{k_1}{c^4 k_0^2}\mathcal{F}^{-1}[\frac{\sin(\frac{k-k_0}{2})}{(\frac{k-k_0}{2})} \frac{\sin(\frac{k+k_0}{2})}{(\frac{k+k_0}{2})}]\\
=&\frac{1}{c^2}(1-\frac{k_1}{c^2 k_0^2})(1-|\xi|)I_{[-1,1]}(\xi) + \frac{k_1}{c^4 k_0^2}[ (e^{i k_0\xi}I_{[-\frac{1}{2},\frac{1}{2}]}(\xi))* (e^{-i k_0\xi}I_{[-\frac{1}{2},\frac{1}{2}]}(\xi)) ]\\
=&\frac{1}{c^4 k_0^2}[(c^2 k_0^2-k_1)(1-|\xi|)-\frac{k_1}{k_0}\sgn(\xi)\sin{(k_0 \xi)}] I_{[-1,1]}(\xi)
\end{split}
\end{equation}
by direct calculation, where $I_{[a,b]}$ has been used to denote
the indicator function in the corresponding interval.
Utilizing the Convolution Theorem, Eq.~(\ref{eqn4_1}) can be transformed back to real space as:
\begin{eqnarray}
\label{eqn4b_1}
R(\xi)=(m_1*(R^{3/2})_+)(\xi),
\end{eqnarray}
where $*$ will be used hereafter to denote the convolution.

Then, by iterating $R^{(l)} \to ((R^{3/2})_+)^{(l)} \to R^{(l+1)}$ using Eq.~(\ref{eqn4b_1}), in the same spirit as the calculation of~\cite{pego1}
(see also the much earlier similar proposal of~\cite{mertens}),
we obtain yet a new scheme, hereafter termed ``scheme II''.
Next, we'll try to find $S(\xi)$ from Eq.~(\ref{eqn4_2}). Theoretically, one can follow the order $R \to \hat{R} \to \hat{S} \to S$ to find $S(\xi)$. But the singularity points of $\hat{S}$ can induce practical difficulties in
this vein of numerical computation. Instead, we use the relationship between $\hat{S}$ and $\widehat{(R^{3/2})_+}$ \begin{eqnarray}
\label{eqn4b_2}
\hat{S}=-\frac{k_1}{c^4 \nu_1(k^2-k_0^2)}\sinc^2(\frac{k}{2})\hat{(R^{3/2})_+}
\equiv M_2(k)\hat{(R^{3/2})_+}
\end{eqnarray}
to reach $S(\xi)$ since $\widehat{(R^{3/2})_+}$ is already known. Similarly to deriving Eq.~(\ref{eqn4b_1}), we find the inverse Fourier transform of $M_2(k)$ as
\begin{equation}
\begin{split}
\label{eqn_m2}
m_2(\xi)=&\mathcal{F}^{-1}[-\frac{k_1}{c^4\nu_1(k^2-k_0^2)}\sinc^2(\frac{k}{2})]\\
=&\frac{k_1}{c^4\nu_1 k_0^2}\mathcal{F}^{-1}[\frac{\sin^2(\frac{k}{2})}{(\frac{k}{2})^2}-\frac{\sin^2(\frac{k}{2})}{(\frac{k^2-k_0^2}{4})}]\\
=&\frac{k_1}{c^4 \nu_1 k_0^3}[k_0(1-|\xi|)+\sgn(\xi)\sin{(k_0 \xi)}]I_{[-1,1]}(\xi)
\end{split}
\end{equation}
and apply the convolution theorem on Eq.~(\ref{eqn4b_2}), to obtain
\begin{eqnarray}
\label{eqn4b_3}
S(\xi)=(m_2*R^{3/2})_+)(\xi).
\end{eqnarray}

\subsection{Finite Domain with Fourier Series}

As a result of the necessity to explore realistic computations and
observations, it is natural to also consider the problem as restricted on a
finite domain, i.e., for $\xi\in [-L, L]$. Then, we will
express the corresponding functions as Fourier series instead of using the
Fourier transform.
In this case, the bounded nature of our profiles, in conjunction
with the finiteness of the domains guarantees integrability.
 $R$, $S$ and $R^p_+$ can then be expressed
in the form of Fourier series $f(\xi)=\sum_{k=-\infty}^{\infty}f_k e^{i\frac{2\pi}{2L}k\xi}$ where $f_k=\frac{1}{2L}\int_{-L}^{L}f(\xi)e^{-i\frac{2\pi}{2L}k\xi}d\xi$. It can be shown when $f(\xi)$ is (piecewise) smooth on the interval $[-L, L]$,
that its Fourier series will converge to $f$ on $[-L, L]$. Since this condition is not difficult to achieve, we assume $R$, $S$ and $R^p_+$ are such functions. Then Eqs.~(\ref{eqn3_1})-(\ref{eqn3_2}) can be written as:
\begin{eqnarray}
\label{eqn5_1}
(k_1+k_1\nu_1-c^2\nu_1 (\frac{\pi}{L}k)^2)R_k&=&\frac{1}{c^2}(k_1-c^2\nu_1 (\frac{\pi}{L}k)^2)\sinc^2{(\frac{\pi}{2L}k)}(R^{3/2}_+)_k,\\
\label{eqn5_2}
(k_1-c^2\nu_1 (\frac{\pi}{L}k)^2)S_k&=&k_1 R_k.
\end{eqnarray}
where $k\in\mathbb{Z}$.

It is straightforward to observe that Eqs.~(\ref{eqn5_1})-(\ref{eqn5_2}) are just discrete versions of Eqs.~(\ref{eqn4_1})-(\ref{eqn4_2}). However, the fact that $k$ only assumes integer values can greatly facilitate the avoidance of
singularities in these equations and hence the unobstructed performance
of the relevant computations. Nevertheless, there are still
requirements on $R_k$ and $((R^{3/2})_+)_k$ so that Eqs.~(\ref{eqn5_1})-(\ref{eqn5_2}) can be defined for any integer $k$:

(c). If $\sqrt{\frac{k_1(1+1/\nu_1)L^2}{c^2\pi^2}}=k_0 \frac{L}{\pi} \in\mathbb{Z}$ and $k_0\neq 2nL$, $((R^{3/2})_+)_{k_0\frac{L}{\pi}}=((R^{3/2})_+)_{-k_0\frac{L}{\pi}}=0$;

(d). If $\sqrt{\frac{k_1 L^2}{c^2\nu_1\pi^2}}=k_2 \frac{L}{\pi} \in\mathbb{Z}$, $((R^{3/2})_+)_{k_2\frac{L}{\pi}}=((R^{3/2})_+)_{-k_2\frac{L}{\pi}}=0$.

When $k_0 \frac{L}{\pi} \not\in\mathbb{Z}$, Eq.~(\ref{eqn5_1}) can be expressed as $R_k=\frac{(k_1-c^2\nu_1 (\frac{\pi}{L}k)^2)}{c^2(k_1+k_1\nu_1-c^2\nu_1 (\frac{\pi}{L}k)^2)}\sinc^2{(\frac{\pi}{2L}k)}(R^{3/2}_+)_k=\tilde{M_1}_k (R^{3/2}_+)_k$. Since $\tilde{m_1}(\xi)=\sum_{k=-\infty}^{\infty}\tilde{M_1}_k e^{i\frac{2\pi}{2L}k\xi}$ converges and $\frac{1}{2 L}\int_{-L}^{L}f_j e^{ij\frac{2\pi}{2L}y} g_k e^{ik\frac{2\pi}{2L}(\xi-y)} dy = f_j g_k e^{i k \frac{2\pi}{2L}\xi}\delta_{jk}$, the equation $R_k=\tilde{M_1}_k (R^{3/2}_+)_k$ can be rewritten using convolution as:
\begin{eqnarray}
\label{eqn5b_1}
R(\xi)=\frac{1}{2 L}(\tilde{m_1}* (R^{3/2})_+ ) (\xi),
\end{eqnarray}
where in this case the convolution is restricted to the domain $[-L, L]$.

In order to get $S$, we define $\tilde{m_2}(\xi)=\sum_{k=-\infty}^{\infty}\frac{k_1}{c^2(k_1+k_1\nu_1-c^2\nu_1 (\frac{\pi}{L}k)^2)}\sinc^2{(\frac{\pi}{2L}k)} e^{i\frac{2\pi}{2L}k\xi}$ and similarly obtain the equation:
\begin{eqnarray}
\label{eqn5b_2}
S(\xi)=\frac{1}{2 L}(\tilde{m_2}*(R^{3/2})_+ ) (\xi).
\end{eqnarray}
Again, the convolution in the equation is only defined
within $[-L, L]$. The above setting realizes a way of finding $R(\xi)$ and $S(\xi)$ on a finite domain which will hereafter be denoted as ``Scheme III''.

\section{Numerical Results of Schemes}

\subsection{Discussion about scheme II and scheme III}

Scheme I does not involve any extra assumptions, aside from the proposed
existence of traveling waves. Our numerical implementation of
this scheme also employs
a discretization to
identify the relevant structure by means of a fixed point iterative
solution of the associated boundary value problem. Thus, it appears to be
the one with the least amount of additional assumptions (cf. the
discussions above) and as such perhaps the one most likely to
converge to the desired solutions. However, a complication here
involves the potentially non-vanishing boundary conditions on a finite domain
and the identification, a priori, of a suitable initial guess that may
properly capture the behavior at $\xi \rightarrow \pm \infty$.

Scheme II, as indicated above, is appropriate only with the antiresonance condition $k_0=2n\pi$. Under this condition, we start the algorithm with a triangle function and find it to converge to a solution of $R$ without oscillating tails. The other assumption $\hat{R}(\pm k_2)=0$ can also be confirmed in the numerical solution of $R$.
It is worth noting that although this scheme is derived in the infinite domain
setting, all the computations associated with it will be realized,
by necessity, on finite domains.

In scheme III,  $k_0$ can assume any value except for ones such that
$k_0 \frac{L}{\pi}\in\mathbb{Z}$. Since the effect of the domain size on the solution is considered here and $L$ is involved explicitly in the equations, we can always adjust $L$ to make $k_0 \frac{L}{\pi}\not\in\mathbb{Z}$, whatever the values of $\nu_1$, $k_1$ and $c$ are.

With $\frac{k_0}{2\pi}\in\mathbb{Z}$ and $\frac{k_0 L}{\pi}\not\in\mathbb{Z}$, there are cases that can be calculated with both scheme II and scheme III. As numerical results suggest, these two schemes end up with the same solution when $k_0$ is eligible for both of them, naturally demonstrating a strong connection between the two different approaches, which can be explained by the relationship between Fourier transform and Fourier series. We note that when a function $f(\xi)$ has compact support $[-L,L]$,
the coefficients $f_k$ of its Fourier Series
(periodic extension with period $2 L$) are related to the Fourier transform of
that function $\hat{f}(k)$ evaluated at certain points, i.e.
$f_k=\frac{\hat{f}(\frac{k \pi}{L})}{2L}$. Thus it can be shown that $m_1(\xi)=\sum_{k=-\infty}^{\infty}(\frac{1}{2L}\int_{-L}^L m_1(y)e^{-i \frac{2\pi}{2L}k y} dy) e^{i \frac{2\pi}{2L}k \xi}=\sum_{k=-\infty}^{\infty}\frac{1}{2L}M_1(k\frac{\pi}{L})e^{i \frac{2\pi}{2L}k \xi}=\sum_{k=-\infty}^{\infty}\frac{1}{2L}\tilde{M_1}_ke^{i \frac{2\pi}{2L}k \xi}=\frac{1}{2L}\tilde{m_1}(\xi)$, implying Eq.~(\ref{eqn4b_1}) and Eq.~(\ref{eqn5b_1}) are two equivalent formulations when both are applicable. Similar arguments can be applied to show the connection between $m_2=\frac{1}{2L}\tilde{m_2}$ and the equivalence of Eq.~(\ref{eqn4b_2}) and Eq.~(\ref{eqn5b_2}) under the same conditions about $L$ and $k_0$.

Besides the antiresonance situation $\frac{k_0}{2 \pi}\in\mathbb{Z}$, direct calculation also reveals that
\begin{equation}
\begin{split}
\label{eqn6}
 &\frac{1}{2L}\tilde{m_1}(\xi)\\
=&\frac{1}{2L}\sum_{k=-\infty}^{\infty}\frac{1}{c^2}(1+\frac{k_1}{c^2}\frac{1}{(k\frac{\pi}{L})^2-k_0^2})\sinc^2(\frac{k\pi}{2L}) e^{i\frac{2\pi}{2L}k\xi}\\
=&\frac{1}{2L}\sum_{k=-\infty}^{\infty}[(\frac{1}{c^2}-\frac{k_1}{k_0^2 c^4})\sinc^2(\frac{k\pi}{2L})-2\frac{k_1}{k_0^2 c^4}\frac{1}{(k\frac{\pi}{L})^2-k_0^2}(\cos(\frac{k\pi}{L})-1+(1-\cos(k))\cos(k_0 L+k\pi))] e^{i\frac{2\pi}{2L}k\xi}\\
=&\frac{1}{2 k_0^2 c^4}[2(c^2 k_0^2-k_1)\max(1 - |x|,0)+k_1(-2|x|\sinc{(k_0x)}+ |x + 1|\sinc{(k_0(x + 1))}+ |1 - x| \sinc{(k_0(1 - x))})]\\
:=&\bar{m_1}(\xi)
\end{split}
\end{equation} holds when $k_0\frac{L}{\pi}+\frac{1}{2}\in\mathbb{Z}$. Similarly it can be shown that $\frac{1}{2L}\tilde{m_2}(\xi)=\frac{k_1}{2 \nu_1 k_0^2 c^4}[2\max(1 - |x|,0)+ 2|x|\sinc{(k_0x)}- |x + 1|\sinc{(k_0(x + 1))}- |1 - x| \sinc{(k_0(1 - x))})]:=\bar{m_2}(\xi)$ under the same condition.
However, there does not exist any explicit function form for the limit of $\tilde{m_i}$ $(i=1,2)$ as $L\to\infty$ since $\tilde{m_i}$ will encounter singularity points at $L=\frac{n\pi}{k_0}$ where $n\in\mathbb{Z}$.

Thanks to the use of the convolution theorem, implementation of scheme II and scheme III has become straightforward, provided the respective constraint
conditions discussed above are
applicable.

\begin{figure}[!htbp]
\begin{tabular}{cc}
\includegraphics[width=8cm]{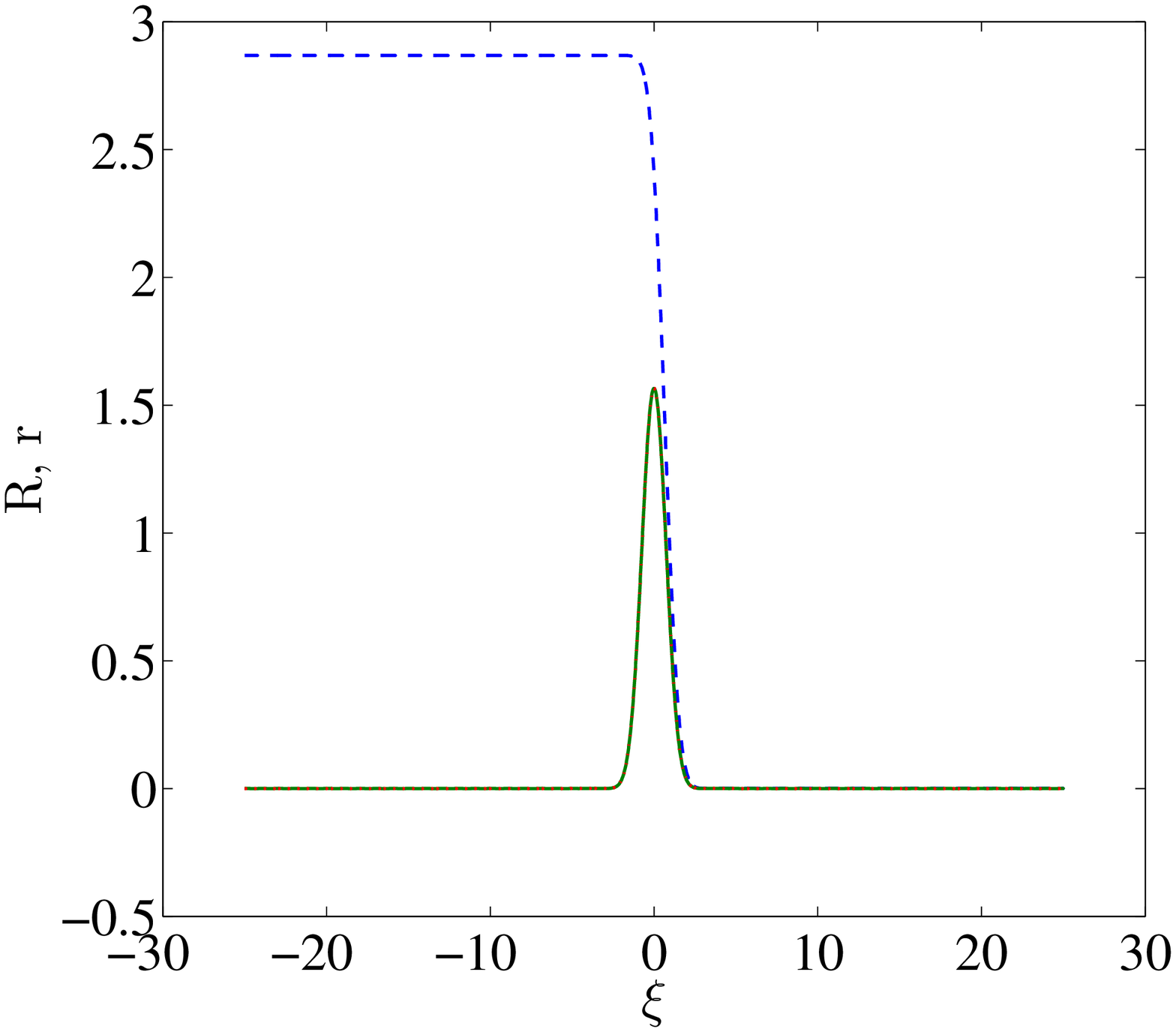}
\includegraphics[width=8cm]{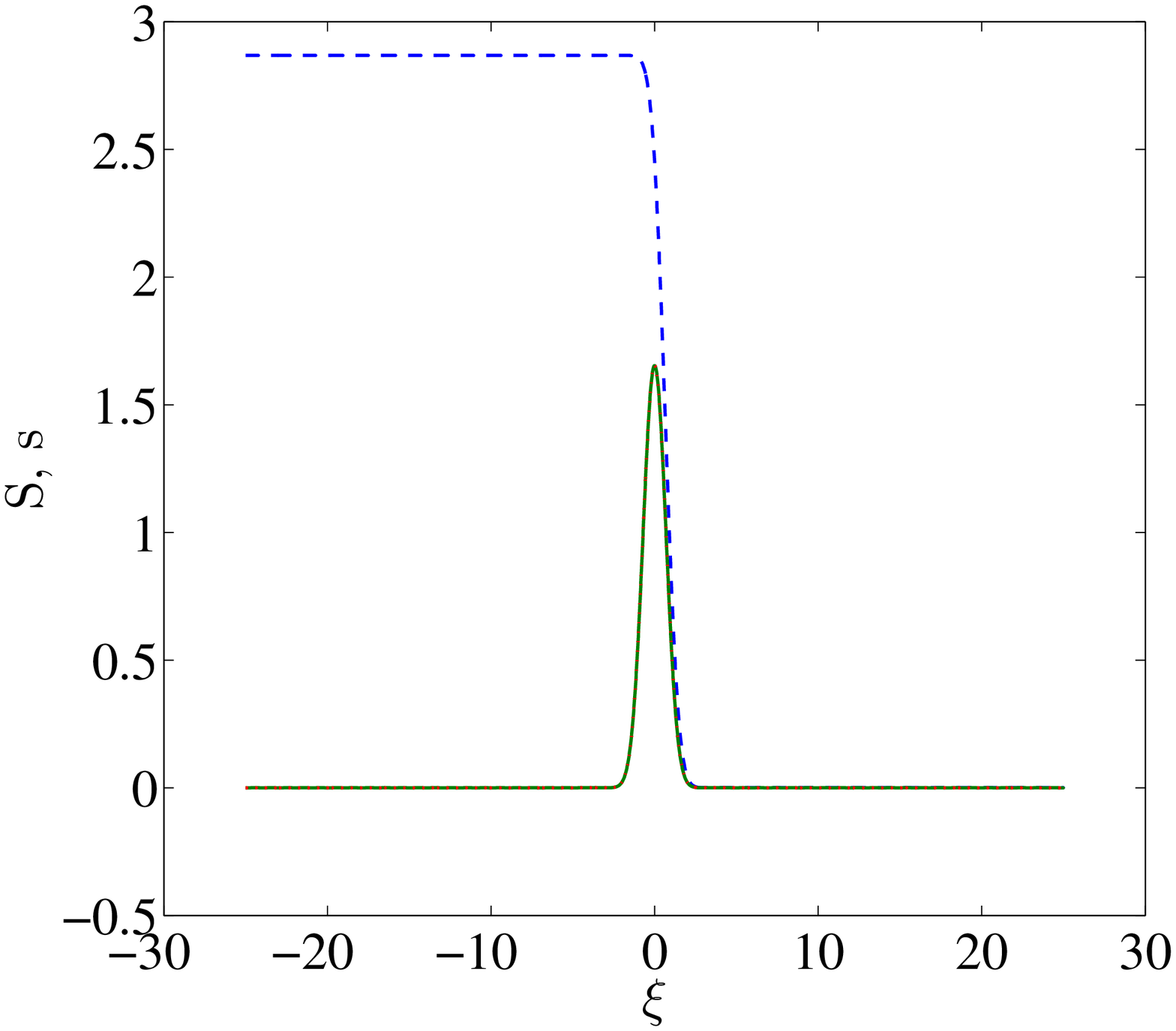}
\end{tabular}
\caption{The left (right) panels showcase the solutions of $R$ ($S$) from scheme I~(solid line) and scheme III~(dotted line) for $k_0=2\pi$ [these results
are effectively identical]. Also, the corresponding solution in the
case of displacements $r$ ($s$) by scheme I (dashed line) is obtained
if we assume $r=s=0$ at the right end of the domain. If we choose $k_0\neq 2n\pi$, the plots will be similar except that all of these solutions will have oscillating tails. It can be seen that the solutions from scheme I and scheme III are indistinguishable to the eye, which confirms the reliability of the Fourier approach. Here we set $c=k_1=1$.}
\label{fg_3}
\end{figure}

\begin{figure}[!htbp]
\begin{tabular}{cc}
\includegraphics[width=8cm]{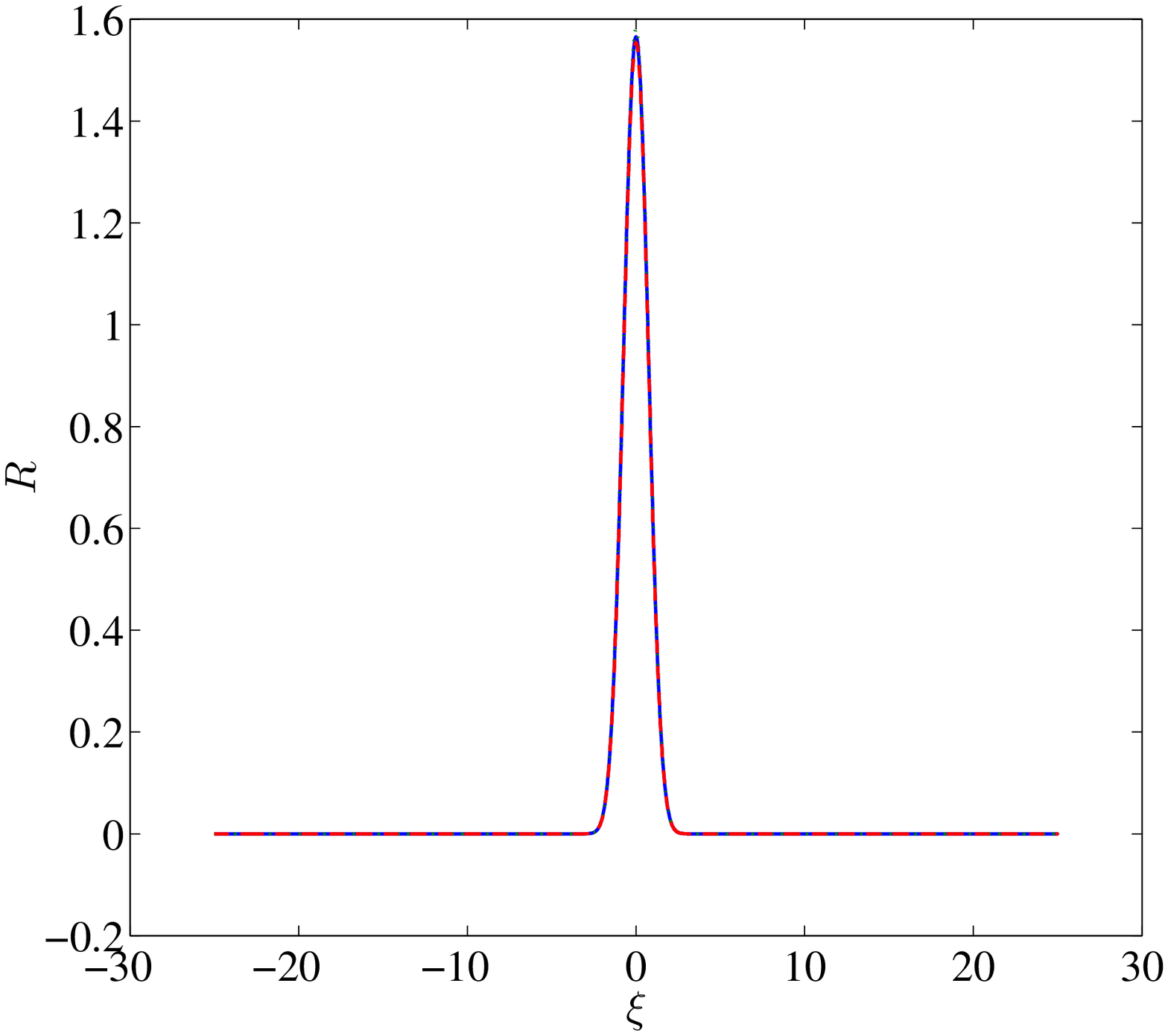}
\includegraphics[width=8cm]{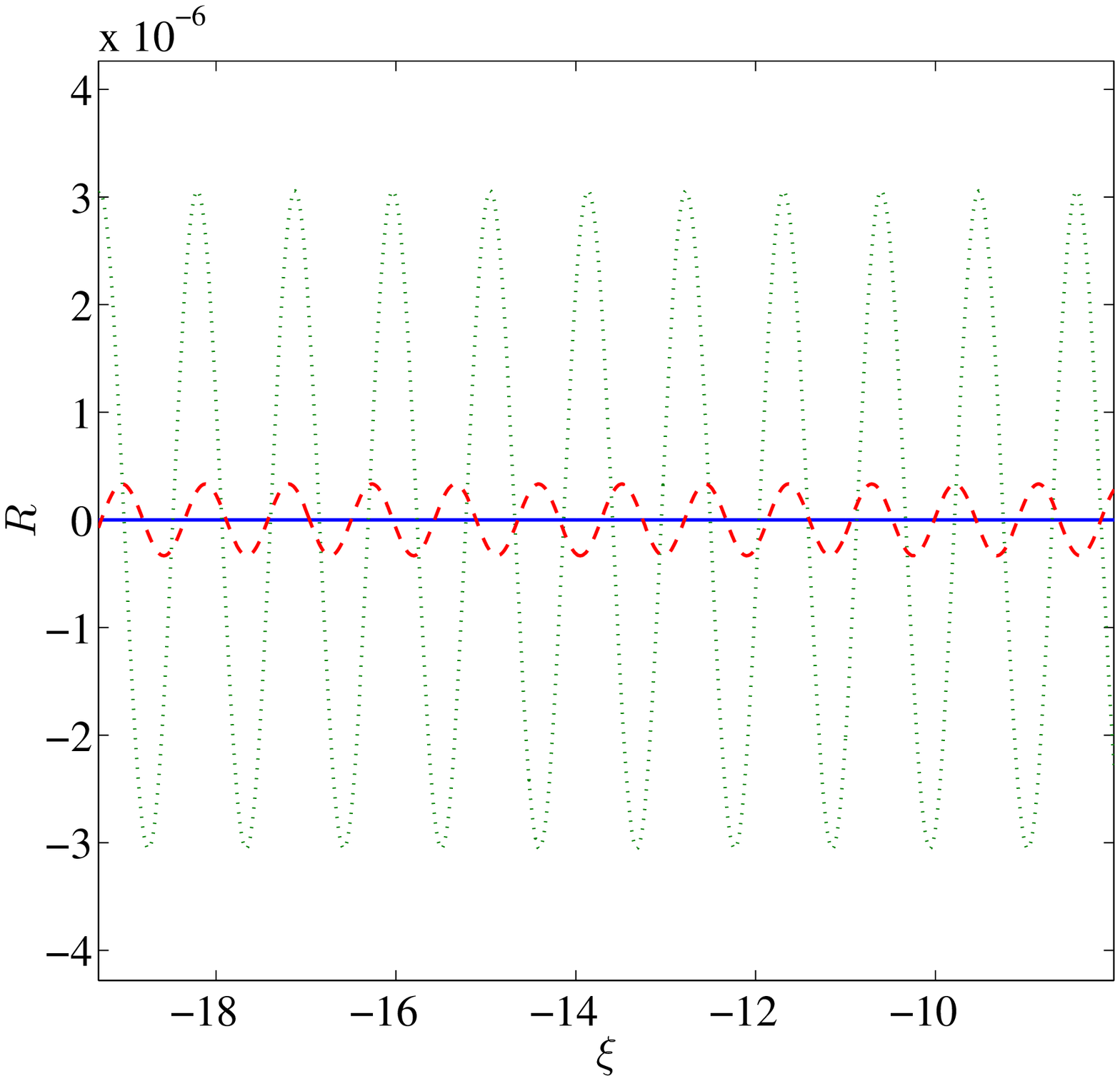}\\
\includegraphics[width=8cm]{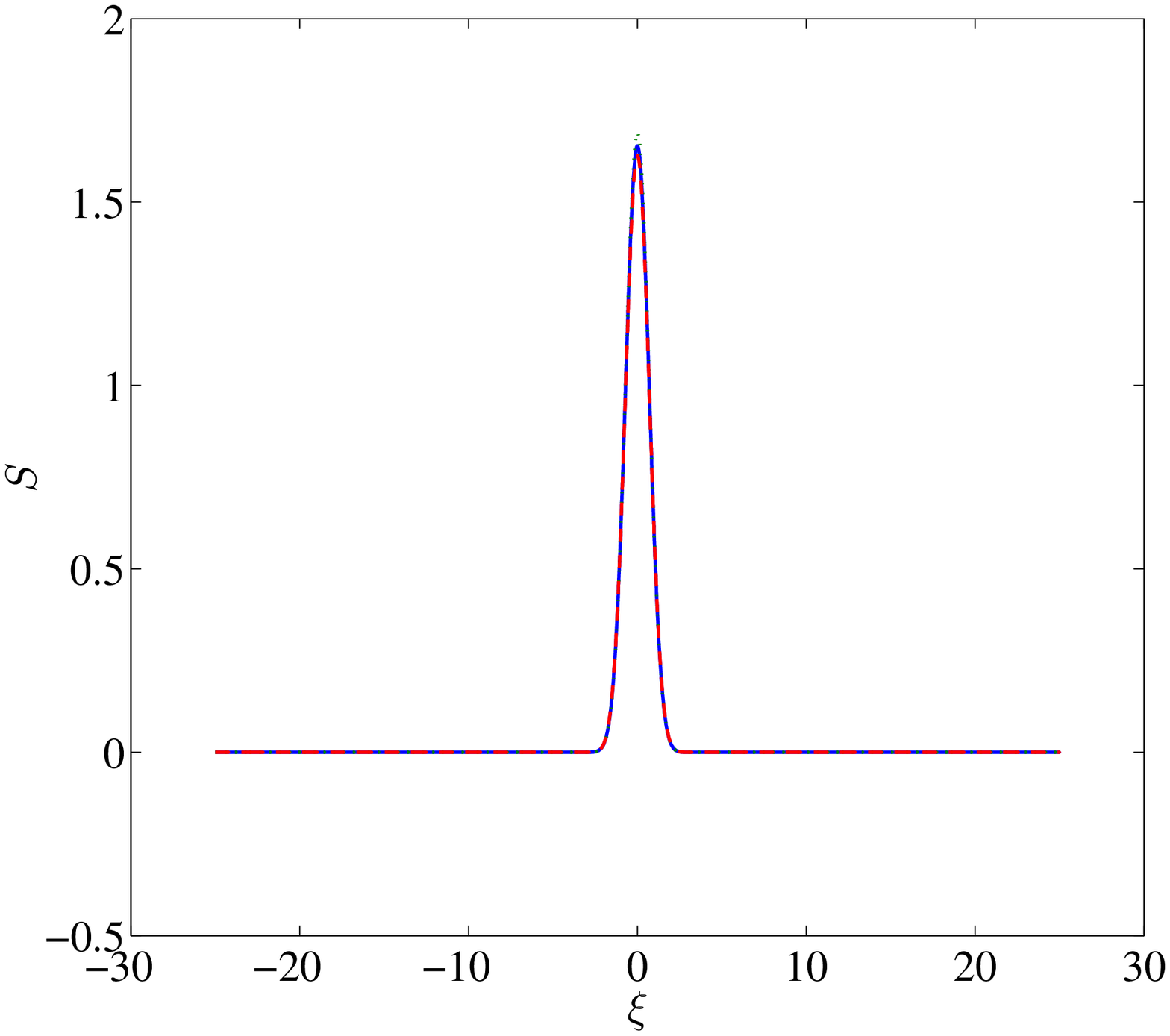}
\includegraphics[width=8cm]{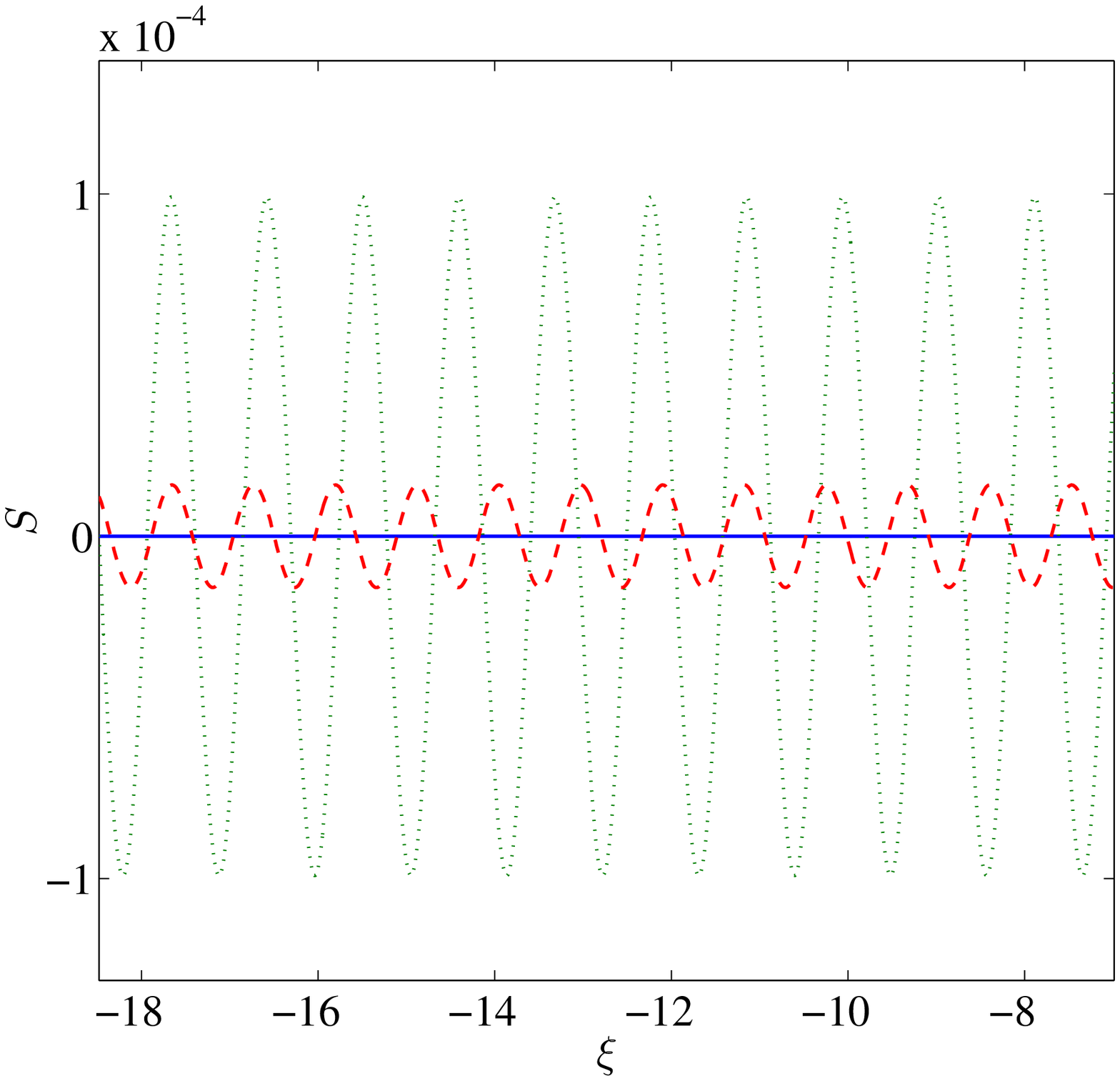}
\end{tabular}
\caption{The top (resp. bottom) panels show the the solutions for $R(\xi)$ (resp. $S(\xi)$) obtained from scheme III for $k_0=2\pi-0.5$ (dashed line), $2\pi$
(solid line), $2\pi+0.5$ (dotted line). The right panels are the corresponding
zooms of the oscillating tails of the solutions in the left panels. Our
computations indicate
that when $k_0\neq2 n \pi$ where $n\in\mathbb{Z}$, the oscillating tails
are generically present.
For these computations, $L=30.16$ and $c=1=k_1=1$.}
\label{III_3}
\end{figure}

If so, our numerical computations of scheme III indicate that
the oscillating tails are present except when $k_0=2 n \pi$
with $n\in\mathbb{Z}$.  In those cases, as shown in Fig.~\ref{fg_3}
(including also computations from Scheme I),
the oscillation tails are absent, while if $k_0 \neq 2 n \pi$, as in
the cases of Fig.~\ref{III_3}, they are generically present. Later in the subsection C, we would discuss more about the different schemes
and their numerical results as a function of the system's parameters.

\subsection{Further discussion about Scheme I}

Scheme III discussed above represents the most direct way of
obtaining a numerically exact (up to a prescribed tolerance) solution
to the traveling wave problem, and typically it constitutes
our most direct technique, as illustrated e.g. in Figs.~\ref{fg_3}-\ref{III_3}.
Additionally, however, a few iterations of this Scheme could
be used to provide us with a good initial guess for
attacking the problem by means of the more direct
Eqs.~(\ref{eqn2_1})-(\ref{eqn2_2}) or Eqs.~(\ref{eqn3_1})-(\ref{eqn3_2}).
In that vein, we find that
upon defining $f=-c^2\ddot{R}(\xi)+(\delta_0+R(\xi+1))_+^p+(\delta_0+R(\xi-1))_+^p-2(\delta_0+R(\xi))_+^p-k_1(R(\xi)-S(\xi))$ and $g=-c^2\nu_1\ddot{S}(\xi)-k_1(S(\xi)-R(\xi))$ where $R$ and $S$ are results of iterations
obtained by scheme III, then the $f$ and $g$ are generally
close to zero.
This confirms that the schemes using Fourier analysis and
the convolution theorem actually provide
solutions
 satisfying, up to a small residual (presumably
created by the discretization) Eqs.~(\ref{eqn3_1})-(\ref{eqn3_2}).

We subsequently tried solving Eqs.~(\ref{eqn3_1})-(\ref{eqn3_2})
using Newton's method, utilizing the solution from scheme III as a
good initial seed for our iterations and as a means for obtaining
information about boundary conditions; i.e., this approach
side-steps both concerns originally present in the context of
the direct method of Scheme I. As indicated by the above residuals,
while the initial guess does not directly solve our system to the
prescribed accuracy,
it is found to be close enough that the Newton will generically,
within our computations,
retain the relevant profile, rapidly converging to a solution of Scheme I,
as shown in Fig.~\ref{fg_3}.
Importantly, we have tested that ``distilling'' this solution and
its time derivative on the lattice (i.e., returning from the variable
$\xi$ to the integer index $n$), we retrieve genuinely traveling
solutions of the original system of differential equations.

\begin{figure}[!htbp]
\begin{tabular}{cc}
\includegraphics[width=8cm]{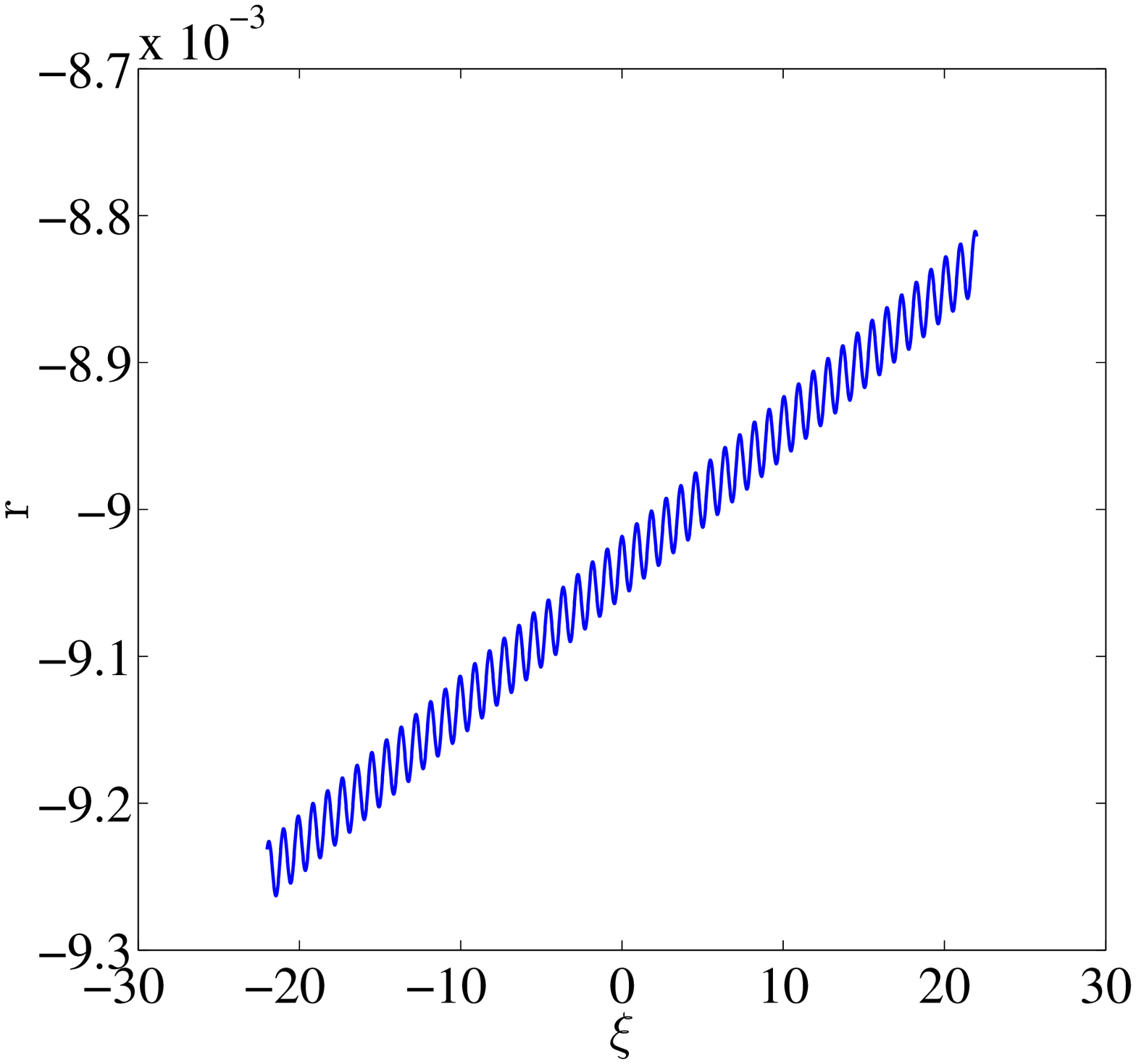}
\includegraphics[width=8cm]{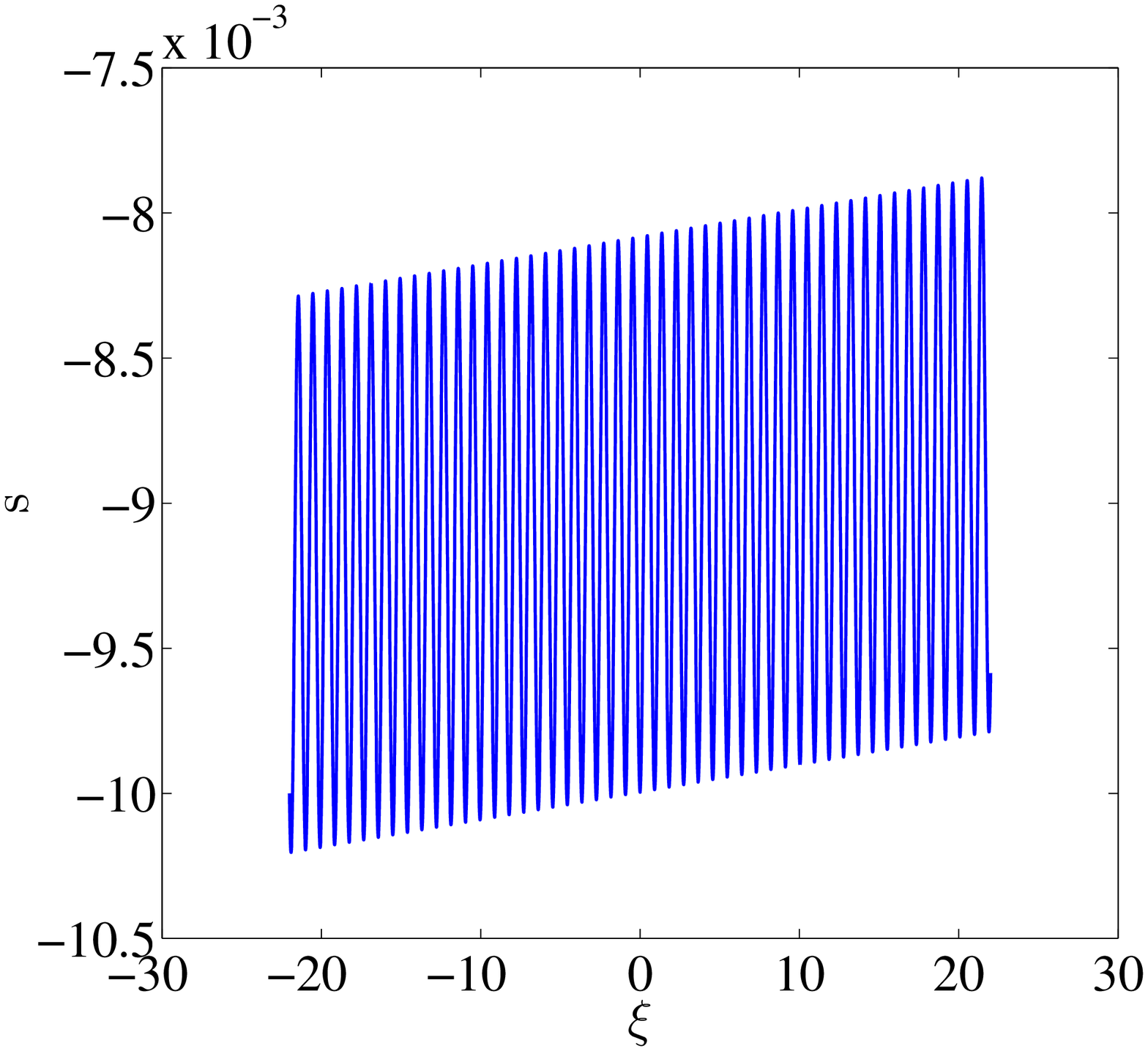}
\end{tabular}
\caption{The top left (right) panel shows another possible, yet less
physically interesting, solution of $r$ ($s$) obtained from scheme I. Here we set $c=k_1=1$, $k_0=2\pi+0.6$.}
\label{fg_4}
\end{figure}

The above schemes provide us with a solution of the strain formulation
problem for $R(\xi)$ and $S(\xi)$.  However, an intriguing
question concerns the reconstruction on the basis of these
fields of the corresponding displacement ones $r$ and
$s$ since $R(\xi)=r(\xi-1)-r(\xi)$ and $S(\xi)=s(\xi-1)-s(\xi)$.
Assuming that we know $r(\xi_0)$, then $r(\xi_0-k)=r(\xi_0)+R(\xi_0-1)+R(\xi_0-2)+...+R(\xi_0-k)$. So in order to fully restore $r$ and $s$, we have to
know their values in an interval of length $1$. Assuming $r$ and $s$ are zero around the right end of the domain, we restored $r$ and $s$ from
$R$ and $S$ in Fig.~\ref{fg_3}, confirming that
they indeed solve Eqs.~(\ref{eqn2_1})(\ref{eqn2_2}).
However, given this ``ambiguity'' in the reconstruction,
it should be noted that $r$ and $s$ obtained in Fig.~\ref{fg_3}
are not the only possible solution pair. For example, the oscillating wave
solutions illustrated in Fig.~\ref{fg_4} can also solve
Eqs.~(\ref{eqn2_1})-(\ref{eqn2_2}). In fact, it has been directly
checked that both the profiles of Fig.~\ref{fg_3}
and those of Fig.~\ref{fg_4} when distilled on the lattice
constitute genuine traveling waves of the original dynamical
problem. Nevertheless, the latter waves are less relevant
for our considerations from a physical perspective, as they
do not constitute fronts at the displacement and pulses at
the strain level.

\begin{figure}[!htbp]
\begin{tabular}{cc}
\includegraphics[width=8cm]{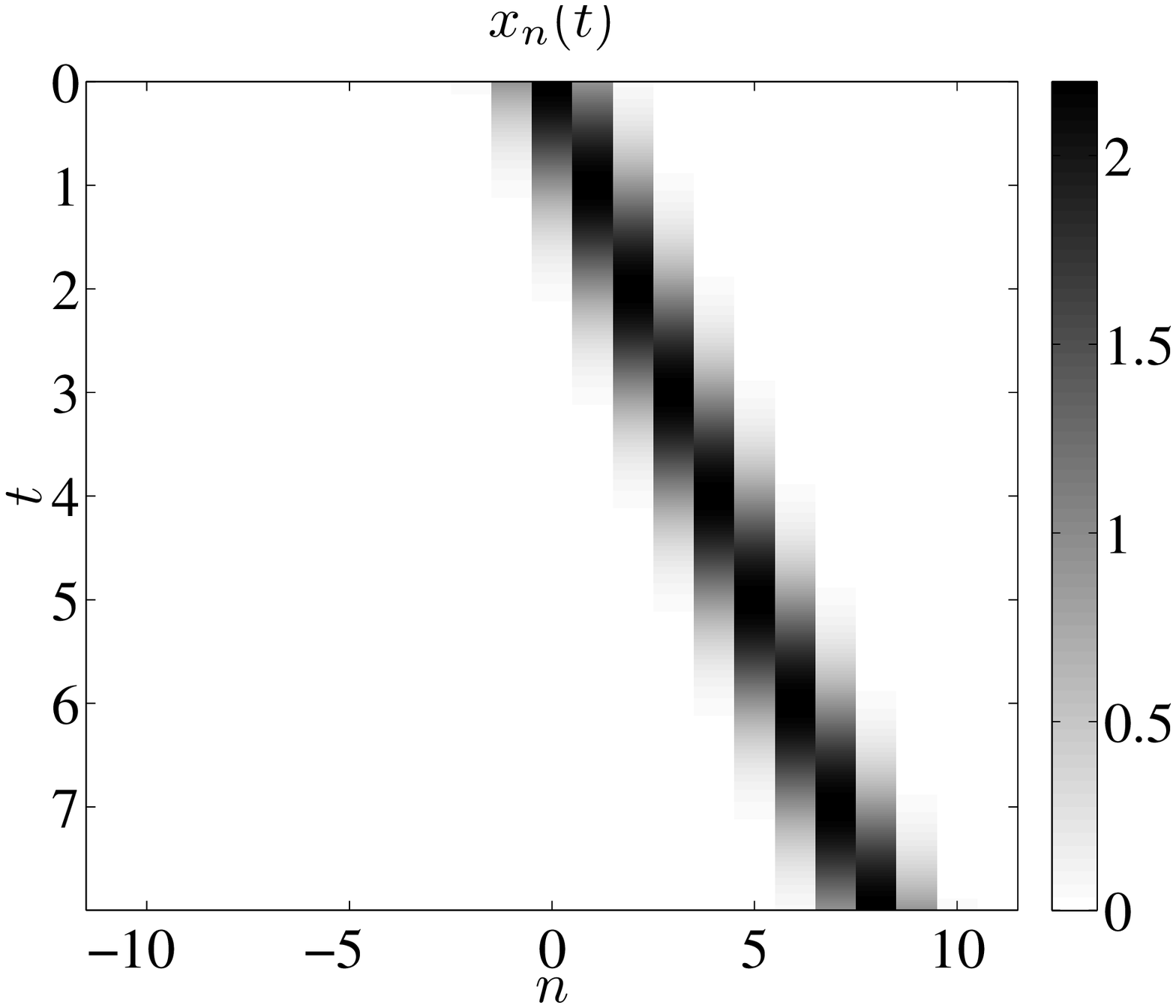}
\includegraphics[width=8cm]{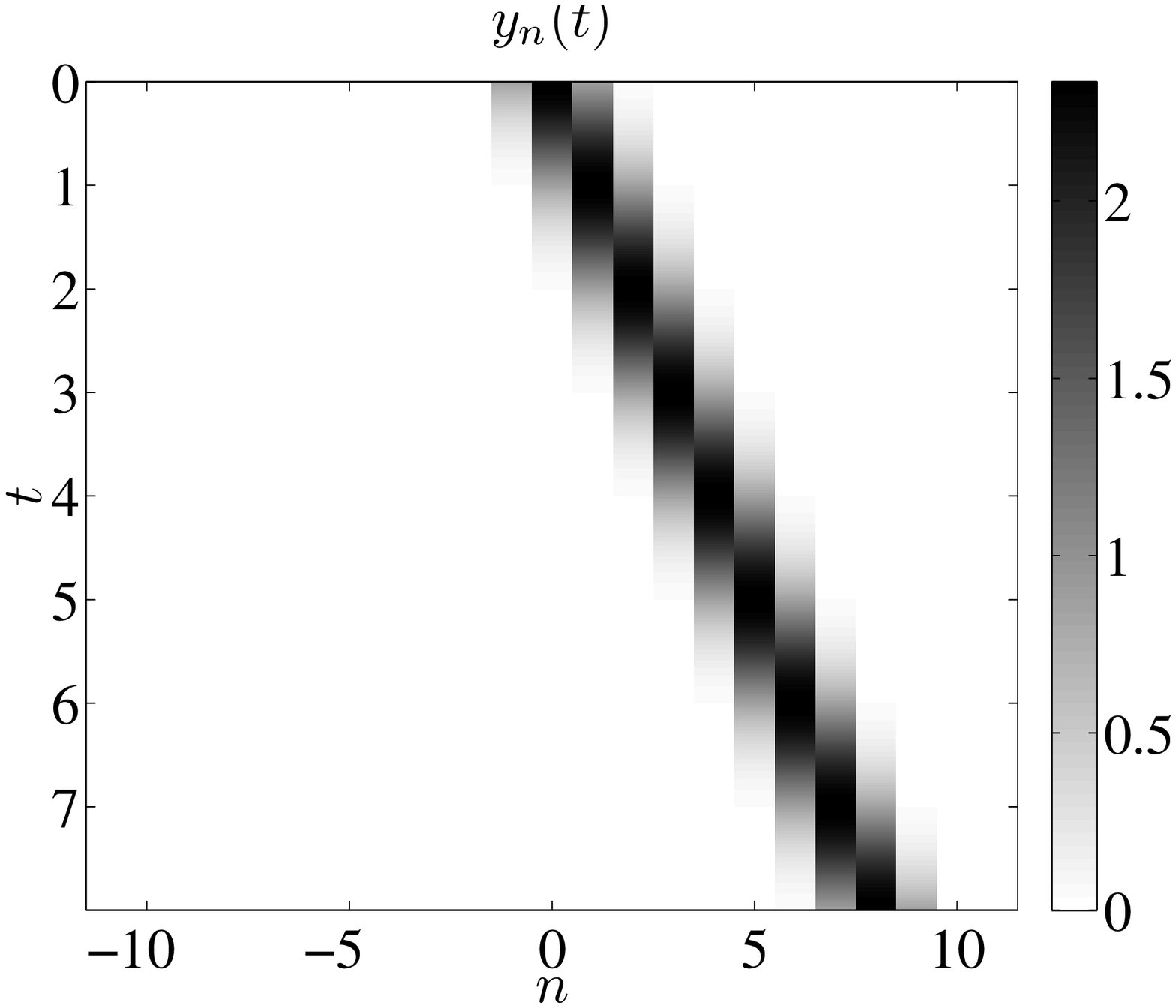}\\
\includegraphics[width=8cm]{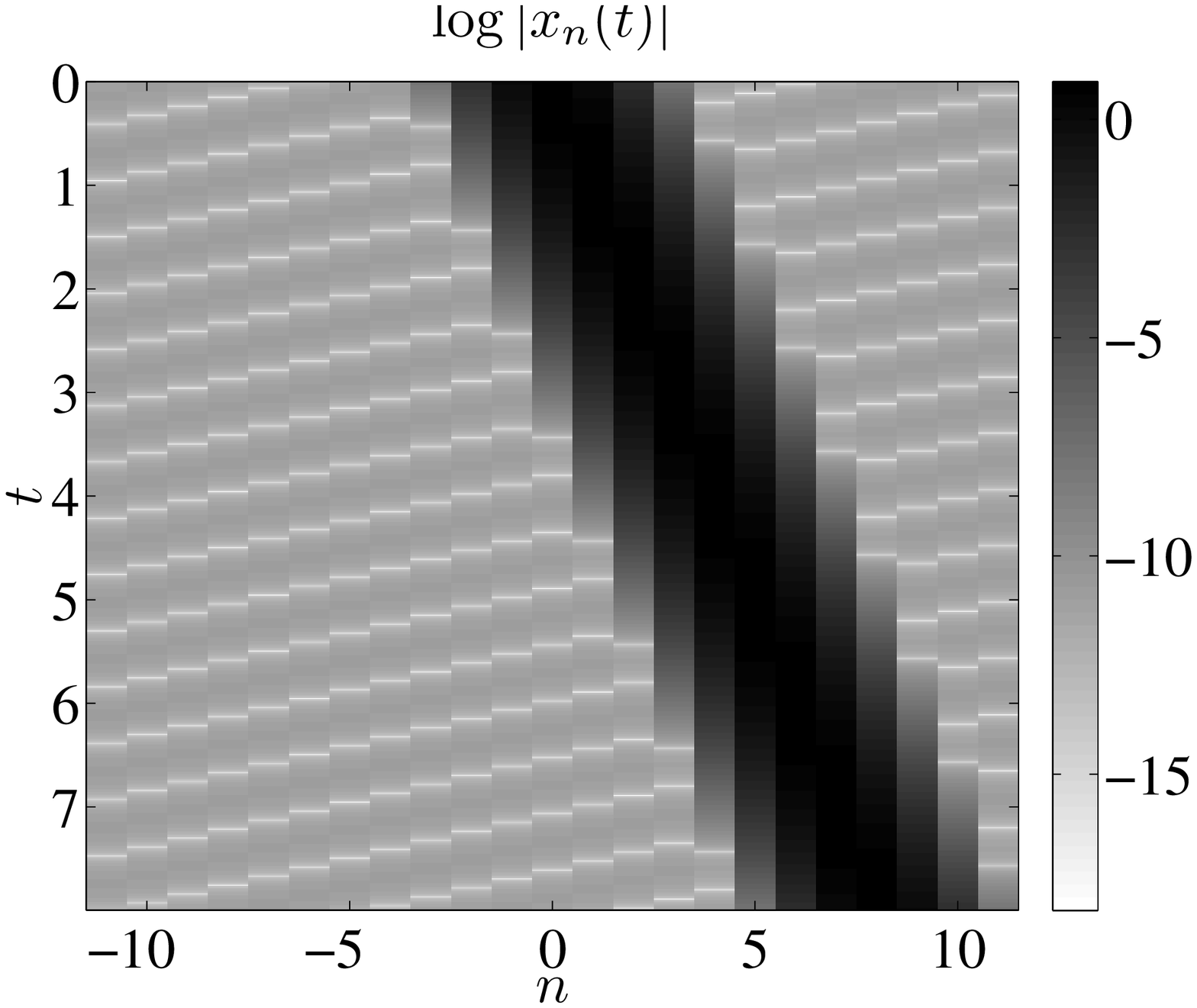}
\includegraphics[width=8cm]{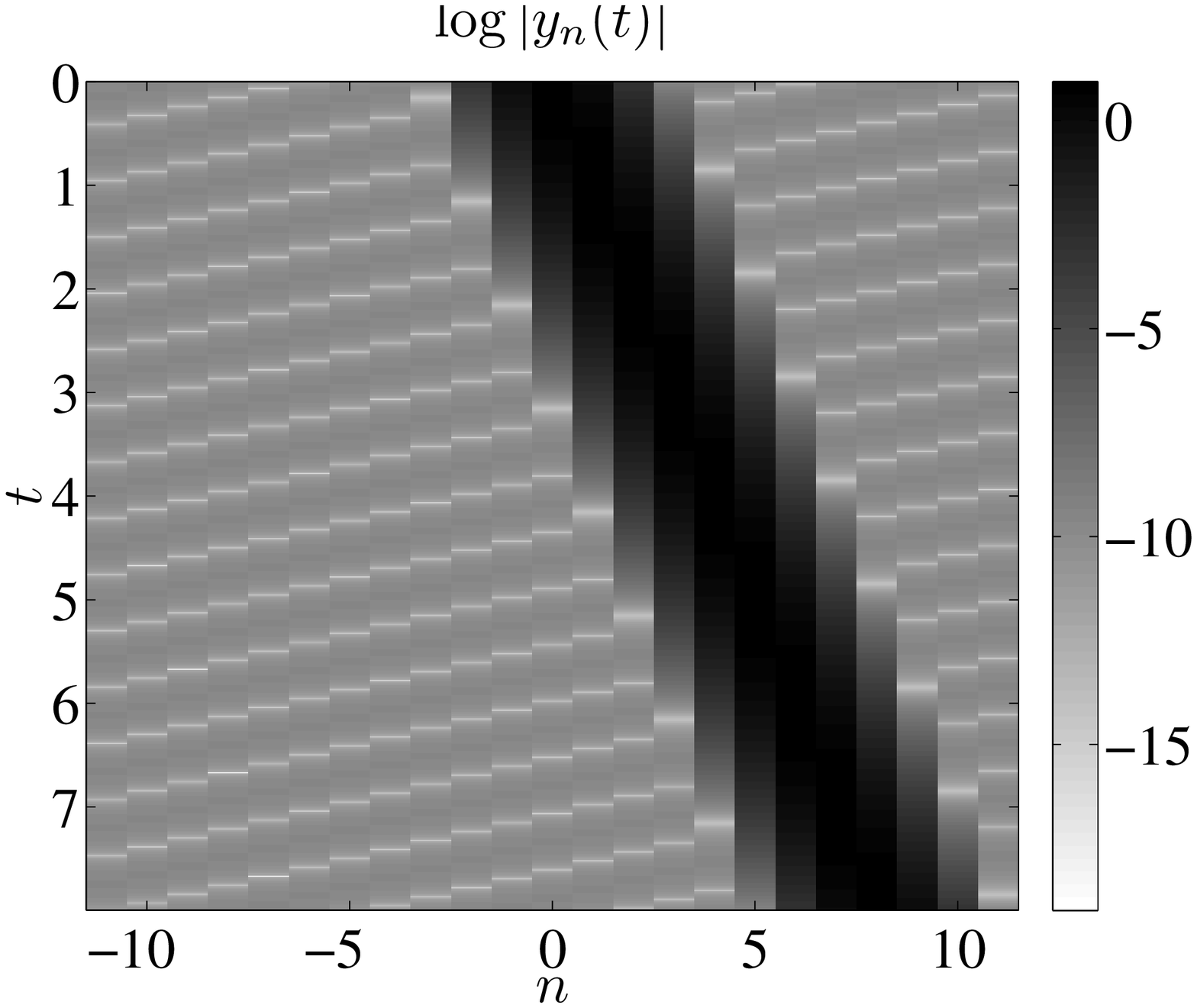}\\
\end{tabular}
\caption{The top panels show the space-time evolution of the traveling wave solution of $x_n$ and $y_n$ to Eqs.~(\ref{eqn3_01})-(\ref{eqn3_02}) for
the antiresonance case $k_0=2\pi$ from Newton's method.
The bottom panels illustrate the corresponding space-time evolutions on
a logarithmic scale
for the case $k_0=2\pi-0.5$. Here, it can be clearly seen that the
oscillation persists in the background.  In these figures,
we have set $c=1$ and $k_1=7$.}
\label{fg_5}
\end{figure}

Lastly, at the level of the present considerations it is relevant
to point out that the identified solutions have been illustrated
in Fig.~\ref{fg_5} to exhibit genuine traveling with the prescribed
speed (of $c=1$). This is done for the anti-resonant case of
$k_0=2 \pi$ (associated with the waveform of Fig.~\ref{fg_3})
in the top panels. Here there is no discernible tail.
It has also, however, been demonstrated in the bottom panel
for the case of
$k_0=2\pi-0.5$. Here, as per Fig.~\ref{III_3}, we expect
the tails to be present, yet they are not observable (due to
their small amplitude, a feature also observed in the experiments
of~\cite{yang}) in a linear scale. For this reason, we have used
a logarithmic scale in the bottom panel of Fig.~\ref{fg_5}, which
clearly showcases the nontrivial traveling oscillatory tail.

\subsection{Study of parameters $k_1$ and $\nu_1$}

Having illustrated how to obtain solutions which are equivalent
between Schemes II and III, and how to utilize these to also
obtain a direct solution from Scheme I, we now turn to the examination
of parametric variations within these schemes. The canonical
parameters whose variations we consider are the linear coupling
with the local resonator $k_1$, as well as the effective
mass $\nu_1$ of the resonator.

When $k_0$ is a multiple of $2\pi$ and gets fixed, consideration of $m_1(\xi)$ proves rather insightful towards understanding
the effects of the parameters $k_1$ and $\nu_1$, since the properties of the solution of Eq.~(\ref{eqn4b_1}) are determined by those of the kernel $m_1(\xi)$.

If $\nu_1$ is fixed and $k_1$ and $c$ vary to retain the same value of $k_0$, $m_1$ will only change in overall amplitude and a family of solutions differing in amplitudes and speeds will be obtained, as we mentioned above in section II.A.
Suppose we fix $k_1$ and change $\nu_1$ and $c$, the shape and the properties of $m_1$ can change greatly. When $c\geq\sqrt{\frac{k_1(1+k_0)}{k_0^2}}$, $m_1(\xi)$ is increasing on $(-\infty,0)$ (hence decreasing on $(0, \infty)$) and always nonnegative. There also exists $c_0\in[0,\sqrt{\frac{k_1(1+k_0)}{k_0^2}}]$ such that $m_1(\xi)$ is nonnegative on $(-\infty,\infty)$ if and only if $c\geq c_0$. Though not necessary, being nonnegative and increasing on $(-\infty,0)$ are
properties of the  kernel that in our numerical computations appear
to facilitate the convergence of scheme II towards a solution. If $k_1$ and $\nu_1$ vary at the same time and $c$ remains unchanged, it is equivalent to considering the previous two cases together and we don't include the
corresponding details here. If we choose $k_0\neq 2n\pi$ but still assign
it a constant value, we will discuss the effects of $\nu_1$ and $k_1$ on the solution of $R$ based on $\tilde{m}_1$ and $\tilde{M_1}_k$ rather than $m_1$ since scheme II becomes inapplicable in this case. Though the period of the oscillating tails is always $\frac{2\pi}{k_0}$ since $k_0$ is fixed, the amplitude of the tails can be very sensitive to other parameters. First, only changing $k_1$ and $c$ to keep the value of $k_0$ fixed can generate a family of solutions with same shape but different amplitudes, just as described in section II.A. When $c$ is fixed and $k_1$ and $\nu_1$ vary, $\tilde{M_1}_k$ is either always increasing or always decreasing over $k_1$. Thus $\max_{\xi}|\tilde{m_1}(\xi)|$, or $\tilde{m_1}(0)$, always grows (or decays) as $k_1$ increases. Again, the case of varying $c$ and $\nu_1$ is just the composition of the previous two cases.

\begin{figure}[!htbp]
\begin{tabular}{cc}
\includegraphics[width=8cm]{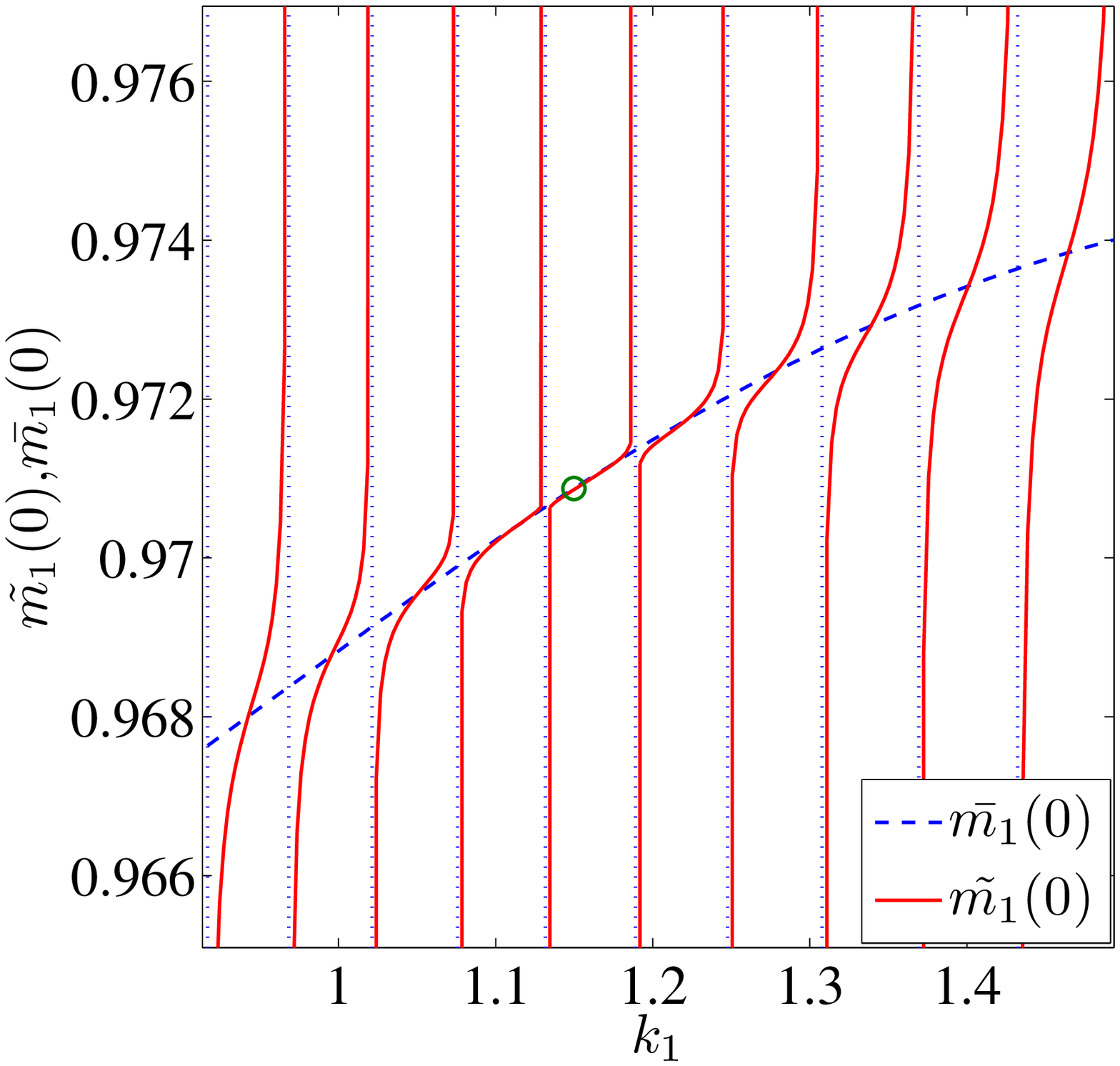}
\includegraphics[width=8cm]{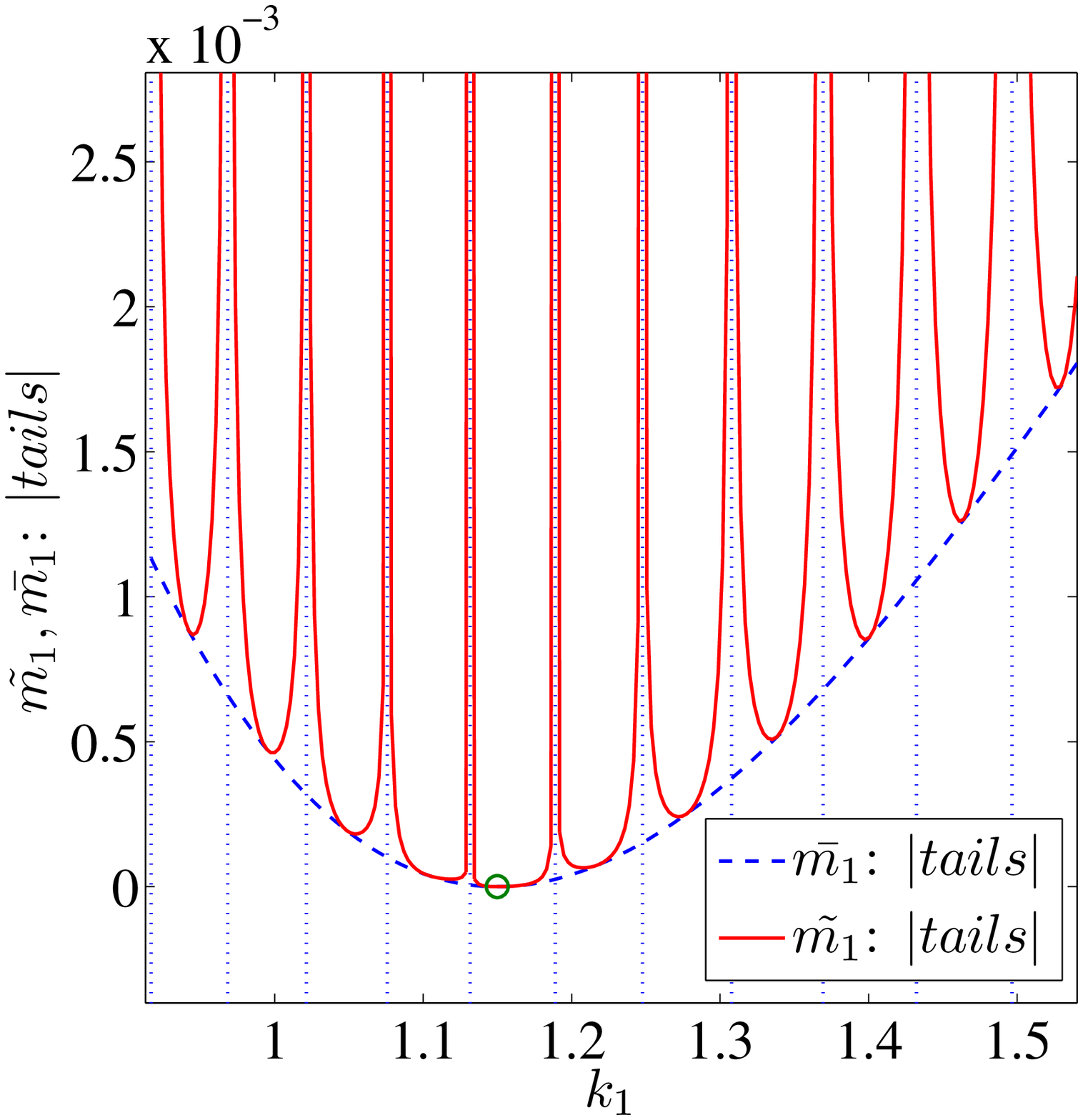}\\
\includegraphics[width=8cm]{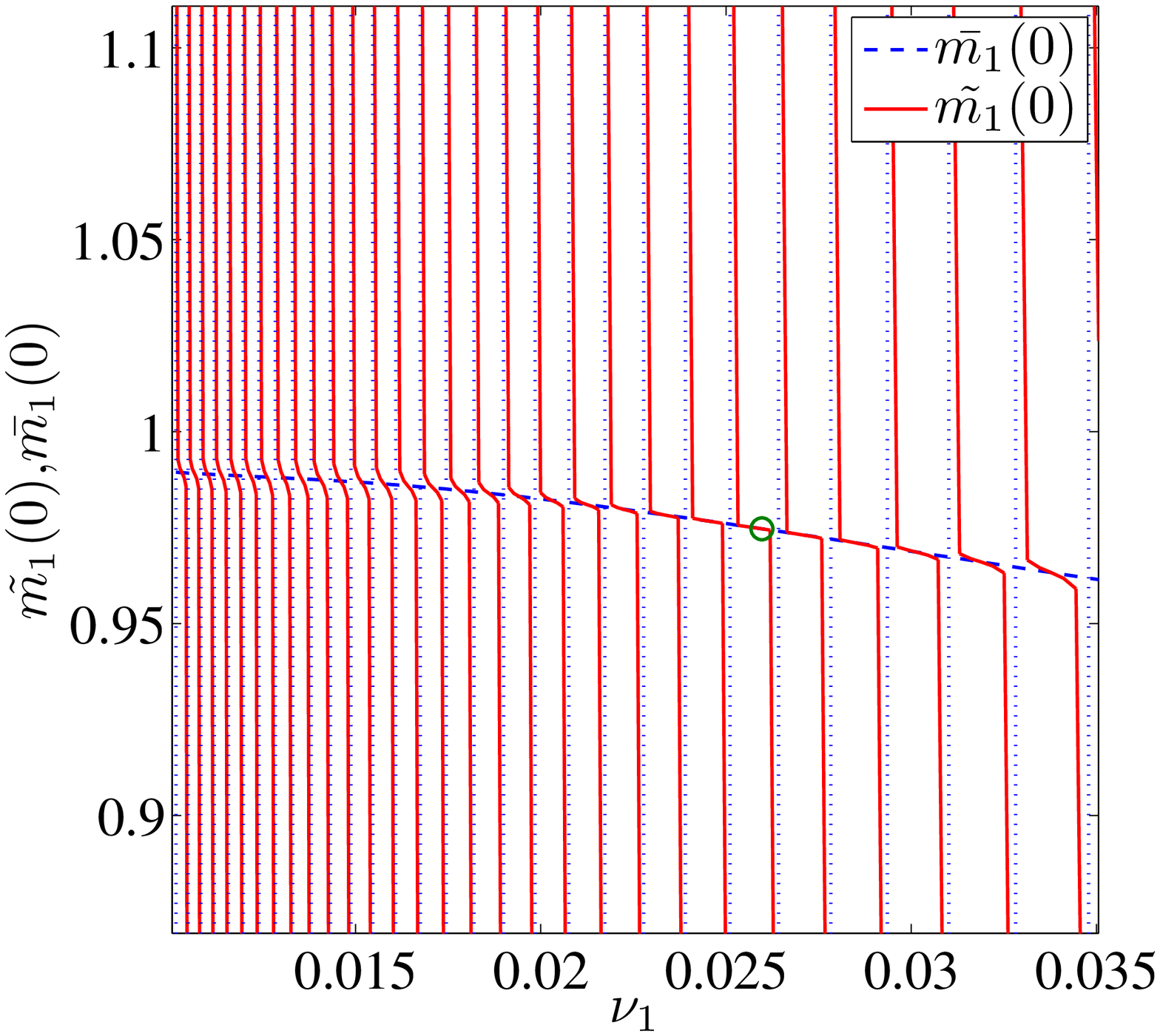}
\includegraphics[width=8cm]{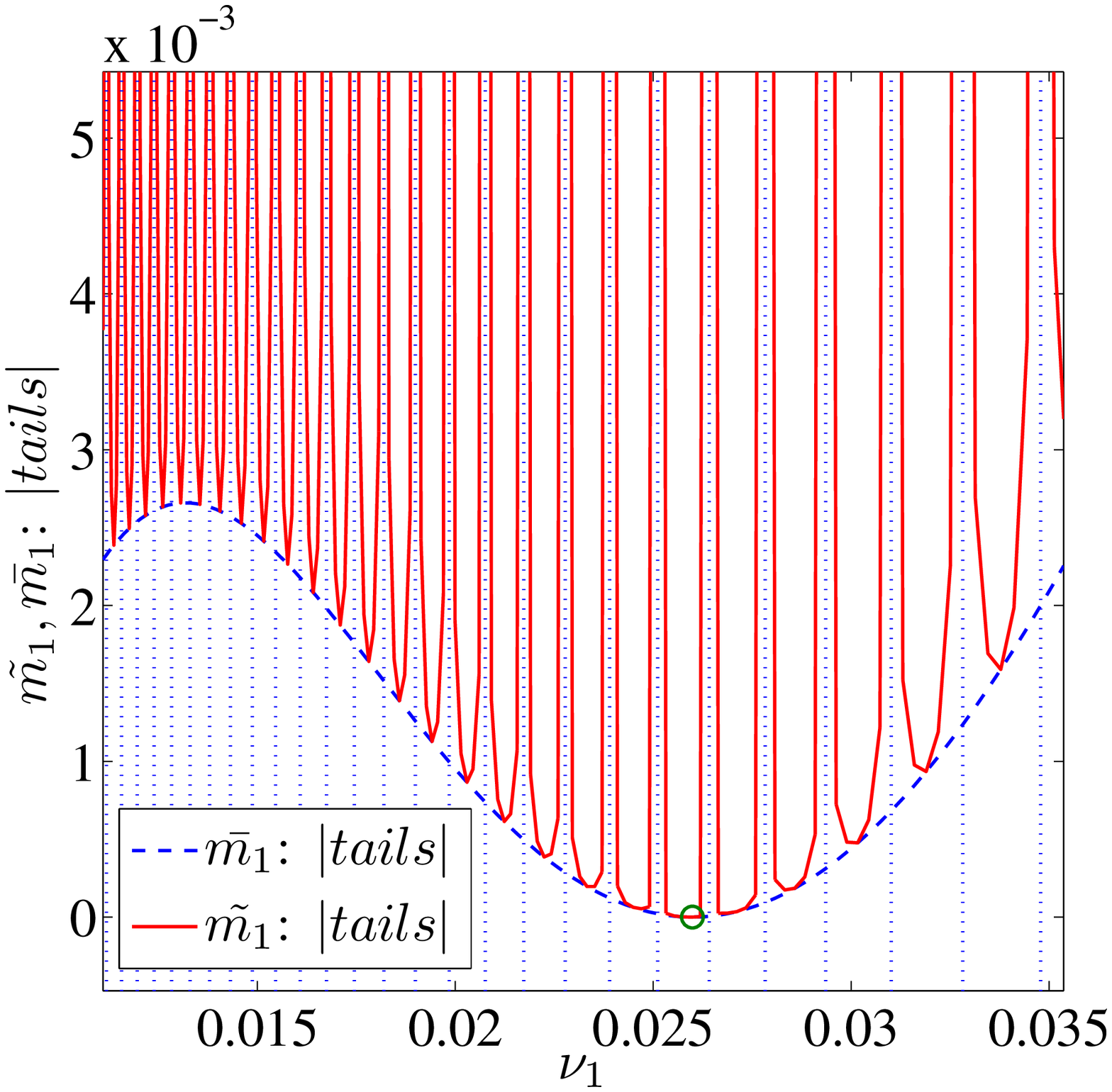}
\end{tabular}
\caption{The top (bottom) left panel shows how $\tilde{m_1}(0)$ (solid line) and $\bar{m_1}(0)$ (dashed line) change over $k_1$ ($\nu_1$) and the top
(bottom) right panel shows the change of the amplitude of oscillating tails of $\tilde{m_1}$ (solid line) and $\bar{m_1}$ (dashed line) as $k_1$ ($\nu_1$) grows. The figures show that $\tilde{m_1}$ blows up every time $k_0\frac{L}{\pi}\in\mathbb{Z}$ but $\tilde{m_1}(0)$ features strictly increasing or decreasing trend between these singularity points (where $k_1$ or $\nu_1$ satisfies $k_0\frac{L}{\pi}\in\mathbb{Z}$). It should be noted that the dotted vertical lines
indicate the singularity points and are added just to better illustrate
the boundaries of each interval. These figures also show $\tilde{m_1}$ intersects with $\bar{m_1}$ when $k_0\frac{L}{\pi}+\frac{1}{2}\in\mathbb{Z}$ or $k_0=2n\pi$ and especially at $k_0=2n\pi$ the kernel $\tilde{m_1}$ loses its oscillating tails (marked by green circles in the right panels). In these computations we set $c=1$, $k_1=1$ and $\nu_1=0.03$ when they are constants. }
\label{tail_1}
\end{figure}

In the discussion above, we assumed $k_0$ as a constant parameter while other
parameters were varied to maintain the constant value of $k_0$. In the following discussion of this subsection, we will allow $k_0$ to vary
and change other parameters freely. If we fix $c$ and $k_1$ and only allow the changing of $\nu_1$, we find that $\tilde{M_1}_k$ for any integer $k$ (hence $\tilde{m_1}(0)$) will decrease
as $\nu_1$ increases from $\frac{k_1}{(\frac{n\pi}{cL})^2-k_1)}$ to $\frac{k_1}{(\frac{(n+1)\pi}{cL})^2-k_1)}$ for any $n\in\mathbb{Z}$. Similarly, we can show that $\tilde{M_1}_k$ increases as $k_1$ increases from $\frac{1}{(1+1/\nu_1)}(\frac{n\pi}{L c})^2$ to $\frac{1}{(1+1/\nu_1)}(\frac{(n+1)\pi}{L c})^2$ or $c^2$ decreases from $(1+1/\nu_1)k_1(\frac{L}{n\pi})^2$ to $(1+1/\nu_1)k_1(\frac{L}{(n+1)\pi})^2$. These strict monotone properties of $\tilde{m_1}$ are revealed in Fig.~\ref{tail_1}. The figure also illustrates the smooth behavior of
$\bar{m_i}$ versus the singular (at the resonance points)
behavor of $\tilde{m_1}$ in the tails
and how these two functions become identical at $k_0\frac{L}{\pi}+\frac{1}{2}\in\mathbb{Z}$.

Due to the nature of the convolution equation $R(\xi)=(m* (R)^{3/2}_+ ) (\xi)=\int_{-L}^L m(y) (R(\xi-y))^{3/2}_+ dy$ with $m$ standing for $\tilde{m_i}$ or $\bar{m_i}$, changes in the kernel $m$ will be translated  into ones for
the solution $R$, as will be explained heuristically below.
Based on our knowledge of $m$ and $R$, we know they are symmetric functions and (arbitrarily splitting them into a core and tail segment)
they can be written as
\begin{eqnarray}
\label{eqn5b_2a}
m(\xi)=m_I(\xi)+m_O(\xi):=m(\xi)I_{[-L_1,L_1]}(\xi)+m(\xi)I_{(-L,-L_1)\cup(L_1,L)}(\xi),\\ R(\xi)=R_I(\xi)+R_O(\xi):=R(\xi)I_{[-L_2,L_2]}(\xi)+R(\xi)I_{(-L,-L_2)\cup(L_2,L)}(\xi)
\end{eqnarray} with $m_I(-L_1)=m_I(L_1)=R_I(-L_2)=R_I(L_2)=0$.
Then the convolution equation can be considered as the sum of $4$ parts $R(\xi)=\int_{-L}^L m(y) (R(\xi-y))^{3/2}_+ dy = A+B+C+D$ where $A:=\int_{-L_1}^{L_1} m_I(y) (R_I(\xi-y))^{3/2}_+dy$, $B:=\int_{-L_1}^{L_1} m_I(y) (R_O(\xi-y))^{3/2}_+ dy$, $C:=\int_{\xi-L_2}^{\xi+L_2} m_O(y) (R_I(\xi-y))^{3/2}_+ dy$ and $D:=\int_{-L}^{L} m_O(y) (R_O(\xi-y))^{3/2}_+ dy$. If we only focus on the cases relevant to our numerical results, we will also assume $\frac{\max_{\xi}|R_I(\xi)|}{\max_{\xi}|R_O(\xi)}\gg\frac{\max_{\xi}|m_I(\xi)|}{\max_{\xi}|m_O(\xi)}\gg\frac{L}{L_1}>\frac{L}{L_2}$.

With all these conditions, it can be shown that the maximum of the solution $\max_{\xi}|R(\xi)|=R(0)=\int_{-L}^L m(y) (R(y))^{3/2}_+ dy=A+B+C+D\approx A=\int_{-L_1}^{L_1} m_I(y) (R_I(y))^{3/2}_+ dy$. As a result, when $m_O$ is changed to $m_{O,new}=q m_O$ (assuming $q=O(1)$ or $q$ such that $m_{new}$ and $I_{new}$ still satisfy our conditions) and $m_I$ is fixed, the maximum of the solution $R_{new}$(namely $R_{new}(0)$) will almost remain at its (previous) value $R(0)$. While if $m_{I,new}=q m_I$ and $m_O$ is unchanged, $R_{I,new}$ will be close to $\frac{1}{q^2}R_I$ thus $R_{new}(0)\approx\frac{1}{q^2}R(0)$. Suppose $\xi_0$ satisfies $R_O(\xi_0)=\max_{\xi}|R_O(\xi)|$ and $L>|\xi_0|>L_1+L_2$, then the amplitude of the tails $R(\xi_0)=A+B+C+D=B+C+D\approx C=\int_{\xi-L_2}^{\xi+L_2} m_O(y) (R_I(\xi-y))^{3/2}_+ dy$ because $A$ becomes 0 and $C$ is dominant over the remaining contributions. By arguments similar to the above, $m_{O,new}=q m_O$ and $m_{I,new}=m_I$ will imply $R_{new}(\xi_0)\approx q R(\xi_0)$. If $m_{O,new}=m_O$ and $m_{I,new}=q m_I$, then $R_{new}(\xi_0)\approx\frac{1}{q^3}R_{new}(\xi_0)$ since $R_{I,new}(\xi)\approx\frac{1}{q^3}R_{I}$.

\begin{figure}[!htbp]
\begin{tabular}{cc}
\includegraphics[width=8cm]{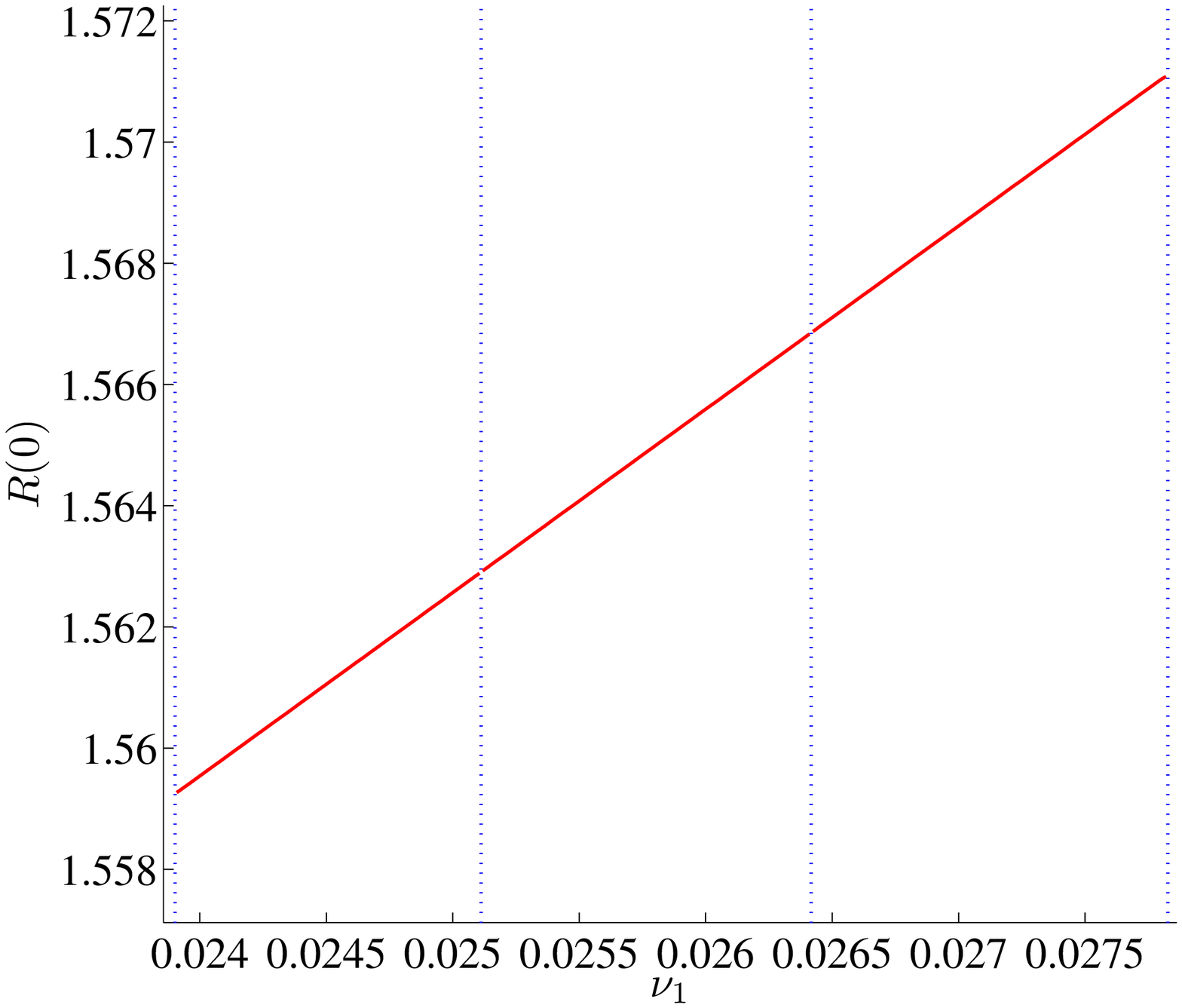}
\includegraphics[width=8cm]{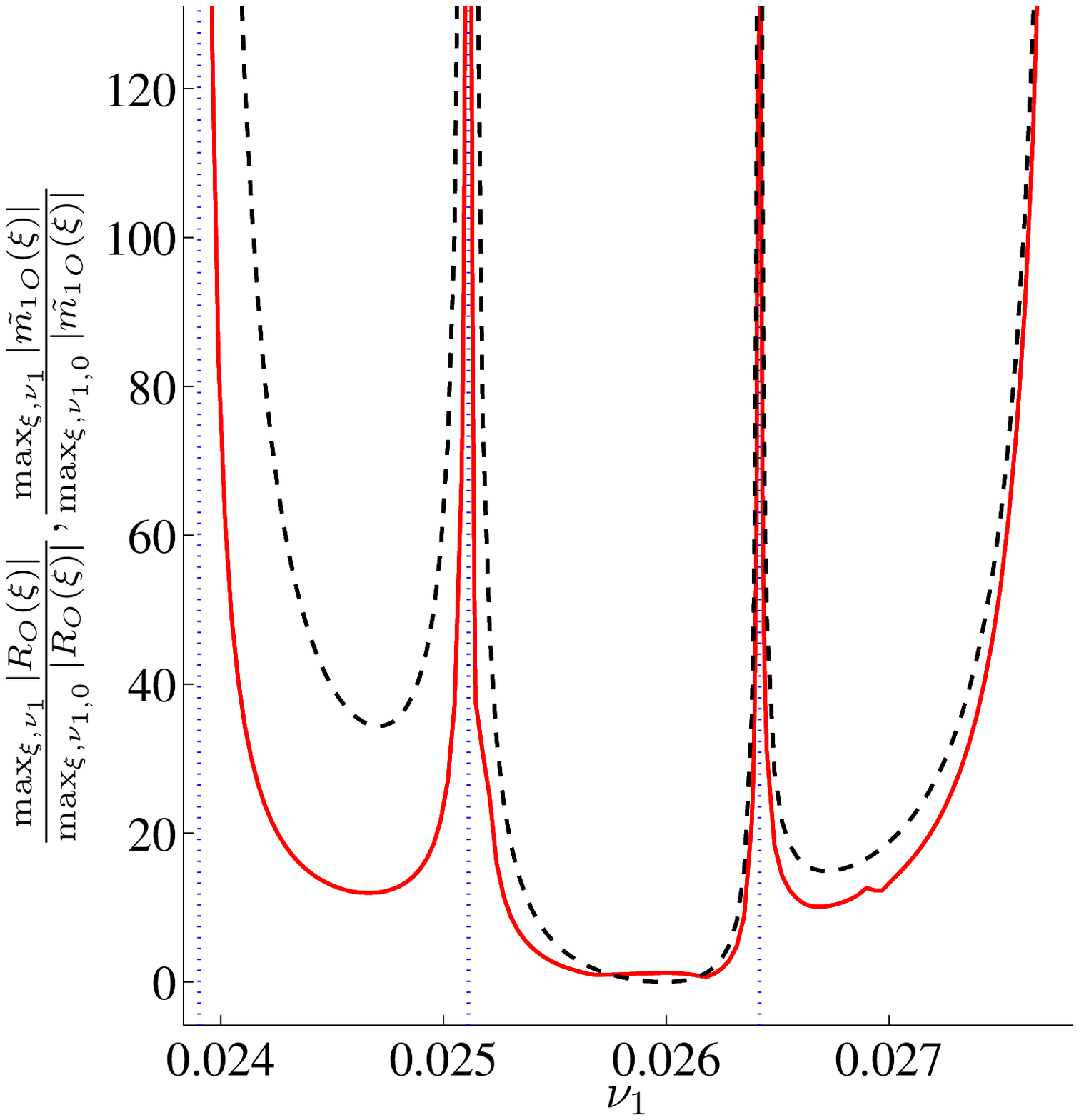}\\
\end{tabular}
\caption{The left panel shows the changing of amplitude of $R$
(or $R(0)$) as $k_1$
grows while the right panel reveals the agreement between the changing of tails of $R$ and that of tails of $\tilde{m_1}$. In the latter panel, the solid red line is for $\frac{\max_{\xi,\nu_1}|R_{O}(\xi)|}{\max_{\xi,\nu_{1,0}}|R_{O}(\xi)|}$ as $\nu_1$ changes and the changing $\frac{\max_{\xi,\nu_1}|\tilde{m_1}_{O}(\xi)|}{\max_{\xi,\nu_{1,0}}|\tilde{m_1}_{O}(\xi)|}$ is described by the black dashed line. Here $\nu_{1,0}=0.02575$, $c=1$ and $k_1=1$ are the parameters we used in the computation.}
\label{tail_2}
\end{figure}

Although the effects of parametric variations on the
kernel $m$, which include changing the shape of $m_I(\xi)$ and the period of $m_O(\xi)$, are much more complicated than merely introducing multiplicative
factors on center or tails, the properties above can still be helpful
towards predicting the changes of $R$ and they are straightforward to apply.
As Fig.~\ref{tail_2} shows, when $\nu_1$ varies in a chosen range and $\tilde{m_1}(0)$ changes slowly, the tails of $\tilde{m_1}$ reflect the form of the
corresponding tails of $R$ very well. At the same time, it should be noticed that $R(0)$ seems not very sensitive about the points $\frac{k_0 L}{\pi}\in\mathbb{Z}$ even though $\tilde{m_1}(0)$ blows up quickly when approaching those points.

\section{Conclusions and Future Challenges}

In the present work, we have revisited a topic of intense
current theoretical, numerical and experimental investigation,
namely the formation of weakly nonlocal traveling waves in
granular chains with local resonators. We attempted to
provide a systematic insight on the different possible
numerical methods that can be used to identify such traveling
waves. Additionally, we highlighted the different challenges
that each of these methods may encounter. In total,
we analyzed three methods. The first
consisted of a direct solution of the
co-traveling wave boundary value problem for the
associated differential advance-delay equation.
The second involved the Fourier representation of
this problem and considered an inverse Fourier transform of the problem to provide an alternative formulation of the
real space problem. This idea was carried out using Fourier transforms defined on the infinite domain. Finally, the third one considered the
Fourier series version of the second
method to extend it to more (and indeed rather
generic) situations. The difficulties existing in each case
were identified and explored, including the limitations
of the Fourier transform (which is restricted to the
anti-resonance case) and of the Fourier series, the need
for a suitable initial guess for the first method
and the boundary conditions thereof, as well as the
issue of converting the strain into a displacement formulation.
We explored how the use of finite domains under
generic non-resonance conditions may enable the
convergence of the third scheme (and how the anti-resonance
enables the convergence of the second scheme to a similar
solution). We then
used good initial guesses from these schemes to
lead to convergent solutions of the first scheme
and were able to reconstruct based on that also
the corresponding displacement profiles.
An interesting byproduct of the parametric variations
considered was the ability to identify anti-resonance parameter
values for which the generic existence of tails
(in our weakly nonlocal solitary waves) was annulled, enabling
the identification of regular, monotonically decaying (on each
side) solitary waves.

While we believe that the above analysis may shed a partial
light on the identification of traveling solitary wave
solutions, there remains a sizable number of open problems
in this direction. Among the significant challenges posed
by the experimental observations of~\cite{yang} we note
the following. For different parameter values than
for the ones where the weakly nonlocal solutions
were identified, it was found that a breathing
while traveling behavior was possible. It would
be extremely interesting to try to identify such
breathing traveling waves and to explore the recurrence
type of dynamics that appears to lead to their formation.
Furthermore, in~\cite{yang}, the nature of experimental excitations
led to the formation of single-sided (i.e., bearing tails
only one side) tails. Exploring the potential of such exact
solutions is an interesting question in its own right.
Another aspect that was briefly explored in~\cite{yang}
was the inclusion of a higher number of resonators.
In the case of more resonators, the possibility
of steady (or even breathing) traveling waves was
found to be less typical. Instead, it was found
that decay of the original pattern's amplitude was
the most commonly observed scenario.
Identifying these cases from the dynamical systems/Fourier
analysis perspective offer presented herein and more
systematically examined the effect of corresponding
parametric variations would constitute a particularly
relevant task for future work.

\bigskip
\textbf{Acknowledgments.}
P.G.K.~acknowledges support from the National Science Foundation under
grant  DMS-1312856, from ERC and FP7-People under grant 605096,
from the US-AFOSR under grant FA9550-12-10332, and from the Binational
(US-Israel) Science Foundation through grant 2010239.
P.G.K.'s work at Los Alamos is supported in part
by the U.S. Department of Energy. A.S. is partially supported by NSF
grant \# 1313107. P.G.K. Also gratefully acknowledges discussions
on this theme with G. Theocharis and D.J. Frantzeskakis.

\end{document}